\documentclass[twocolumn,american]{revtex4-2}
\usepackage[T1]{fontenc}
\usepackage[utf8]{inputenc}
\usepackage{color}
\usepackage{float}
\usepackage{mathrsfs}
\usepackage{bm}
\usepackage{amsmath}
\usepackage{amssymb}
\usepackage{graphicx}

\makeatletter
\setcitestyle{super}

\newcommand{\inlinecite}[1]{Ref.~\citenum{#1}}

\usepackage[hidelinks]{hyperref}

\ifdefined\showcaptionsetup
 \PassOptionsToPackage{caption=false}{subfig}
\fi
\usepackage{subfig}
\makeatother

\usepackage{babel}
\begin{document}
\title{Standard basis operator method for ground-state and temperature properties
of single and two-component Bose-Hubbard model}
\author{Oliwier Urbański}
\email{oliwier1459@gmail.com}
\author{Tomasz P. Polak}
\email{tppolak@amu.edu.pl}
\affiliation{Faculty of Physics and Astronomy, Adam Mickiewicz University of Poznań,
Uniwersytetu Poznańskiego 2, 61-614 Poznań, Poland}
\begin{abstract}
We formulate an improved standard basis operator (SBO) method for
the single and two-component Bose-Hubbard model in three dimensions.
In the first case, nonzero temperature predictions are qualitatively
and quantitatively enhanced by taking into account necessary number
of on-site states, not just three as in previous works. Performance
of the final numerical calculations is also improved by asymptotic
analysis of the self-consistent equations near the critical line.
Obtained results are compared with Monte-Carlo, tensor networks, Quantum
Rotor Approach and experimental data. In the two-component case, SBO
generalizes rather poorly, being able to account for intra-species
thermal and quantum fluctuations, but not the inter-species ones.
Deeper reasons for this situation are discussed. Still however, non-trivial
deformation of the phase diagrams is predicted, together with first-order
phase transitions steered by the changes in chemical potential.
\end{abstract}
\maketitle

\section{Introduction}

The Bose-Hubbard model\citep{gersch1963quantum,fisher1989boson} has
become a benchmark for testing methods for handling strongly-interacting
many-body quantum systems. The mean-field method, known from its ease
of use, provides qualitatively correct predictions in three dimensions
and, at zero temperature, also in two dimensions.\citep{krauth1992gutzwiller,jaksch1998cold,van2001quantum}
However, even in these cases, mean-field results are not satisfactory
quantitatively, as compared to Monte-Carlo results.\citep{capogrosso2008monte,capogrosso2007phase}
A method being a natural extension of the mean-field approach, while
simultaneously providing access to excitation spectra and finite-temperature
observables within the same Green-function formalism, might provide
more accurate results while remaining computationally light-weight.
Among the available semi-analytical techniques, the standard basis
operator formalism occupies an interesting position. While remaining
computationally inexpensive, it provides direct access not only to
the phase boundary but also to single-particle Green’s functions and,
consequently, excitation spectra and momentum-space observables. Unlike
numerically exact approaches, its computational cost remains moderate,
whereas in contrast to perturbative strong-coupling expansions it
is formulated in a nonperturbative Green-function framework. These
features make SBO an attractive candidate for systematic improvements
aimed at increasing quantitative accuracy without sacrificing analytical
transparency.

There is no such an obvious generalization, but a suitable candidate
was provided by Haley and Erdös.\citep{haley1972standard} Their scheme
is based on writing the equation of motion for Green's functions.
Naturally, a lower-order Green's function couples to higher-order
one, which produces an infinite hierarchy. The Random Phase Approximation
(RPA) is used to truncate this ladder by reducing three-operators
Green's functions to two-operators ones. In this sense the method
constitutes a natural next-step beyond mean-field approximation. Calculations
are performed by means of the standard basis operators (SBO), which
allow writing the self-consistency condition.

While the initial usage was for spin-1 Heisenberg ferromagnet,\citep{haley1972standard}
the method was later successfully applied to the Bose-Hubbard model.\citep{sajna2015ground}
Obtained zero-temperature phase diagrams were already very close to
the Monte-Carlo curves. Although the method works also at nonzero
temperatures, its accuracy was worse and, as can be checked, some
artificial discontinuities in the lobes arise. These problems stem
from the three-states approximation -- truncating the Hilbert space
just to three possible single-site states. Since all observables are
reconstructed from the same Green’s functions, inaccuracies originating
from Hilbert-space truncation propagate consistently throughout the
whole finite-temperature description. The importance of this limitation
becomes particularly pronounced at finite temperatures. Thermal population
of higher local Fock states influences not only the position of the
phase boundary but also excitation spectra, momentum distributions
and critical temperatures, all of which are directly accessible within
the Green-function formalism. Consequently, extending the local Hilbert
space is expected to improve several experimentally relevant observables
simultaneously rather than merely refining the critical line.

Another problem with the standard basis operator method is that there
is unwanted freedom in writing the self-consistency equations.\citep{sajna2015ground}
This issue is mitigated by choosing heuristically one combination
and checking its performance against other choices. However, reason
behind the ambiguity has not been discussed yet and it is not clear
how to use the method for more complex models. A natural extension
would be the two-component Bose-Hubbard model, which is already known
to host multitude of interesting phases.\citep{lingua2015demixing,powell2009magnetic,altman2003phase,kuklov2004superfluid,paredes2003cooper,benjamin2014variational}

This work focuses on improving the mentioned issues. Three dimensional
(cubic) lattice is chosen to access Mott-insulator to superfluid phase
transition at nonzero temperatures \citep{mermin1966absence,hohenberg1967existence}.
Section \ref{sec:Formalism} introduces the used formalism and notation
for standard basis operators and Green's functions. Section \ref{sec:Self-consistency-equations}
derives the final self-consistency equations for the Bose-Hubbard
model, extending the previous work\citep{sajna2015ground} by including
an arbitrary number of on-site states and improving numerical performance.
The latter is done by analyzing asymptotic behavior of the mentioned
equations near the critical line. Details are presented in Appendix
\ref{sec:Critical-line-and}. Section \ref{sec:Results-for-single-component}
shows the results for the single-component Bose-Hubbard model, comparing
them with other methods and experiment. Section \ref{sec:SBO-for-two-component}
generalizes Sec. \ref{sec:Self-consistency-equations} for the two-component
case, with resulting plots shown in Sec. \ref{sec:Results-for-two-component}.
Conceptual problems with the SBO method regarding ambiguity in the
self-consistency equations are discussed in Appendix \ref{sec:Conceptual-problems-with}.
Alternative, but unsuccessful, approach to the two-component case
in analyzed in Appendix \ref{sec:Using-a-single}. The paper is ended
with a short conclusion (Sec. \ref{sec:Conclusion}).

\section{Formalism\label{sec:Formalism}}

Suppose we deal with a lattice model, which Hilbert space $\mathscr{H}$
is a tensor product of single-site Hilbert spaces $\mathscr{H}_{i}$,
each labeled by the site index $i$. Often, there exists a natural
(hence the name ``standard'') choice of the basis for each $\mathscr{H}_{i}$.
In our case of Bose-Hubbard models, these are simply Fock states.
Standard basis operator $\hat{L}_{\alpha\beta}^{i}$ is defined as
a transition operator from a state $\beta$ to state $\alpha$ on
site $i$, which can be written as
\begin{equation}
\hat{L}_{\alpha\beta}^{i}\equiv\left(\left|\alpha\right\rangle \left\langle \beta\right|\right)_{i}\equiv I\otimes\cdots\otimes\underbrace{\left|\alpha\right\rangle \left\langle \beta\right|}_{i\text{-th place}}\otimes\cdots\otimes I.\label{SBO def}
\end{equation}

Simple mathematical properties of these operators arise directly from
their definition, which include Hermitian conjugation, identity resolution
and commutation relation:
\begin{equation}
\left(\hat{L}_{\alpha\beta}^{i}\right)^{\dag}=\hat{L}_{\beta\alpha}^{i},\label{Ldag}
\end{equation}
\begin{equation}
\sum_{\alpha}\hat{L}_{\alpha\alpha}^{i}=\mathbb{I},\label{unity_resol}
\end{equation}
\begin{equation}
\left[\hat{L}_{\alpha\alpha^{\prime}}^{i},\hat{L}_{\beta\beta^{\prime}}^{j}\right]=\left(\delta_{\alpha^{\prime}\beta}\hat{L}_{\alpha\beta^{\prime}}^{i}-\delta_{\beta^{\prime}\alpha}\hat{L}_{\beta\alpha^{\prime}}^{i}\right)\delta_{ij}.\label{Lcomm}
\end{equation}
Any single-site operator $\hat{A}^{i}$ can be rewritten in terms
of SBO as
\begin{equation}
\hat{A}^{i}=\sum_{\alpha\alpha^{\prime}}\left\langle \alpha\right|_{i}\hat{A}^{i}\left|\alpha^{\prime}\right\rangle _{i}\hat{L}_{\alpha\alpha^{\prime}}^{i}\label{single-site}
\end{equation}
and any two-site operator as
\begin{equation}
\hat{A}^{ij}=\sum_{\alpha\alpha^{\prime}\beta\beta^{\prime}}\left\langle \alpha\beta\right|_{ij}\hat{A}^{ij}\left|\alpha^{\prime}\beta^{\prime}\right\rangle _{ij}\hat{L}_{\alpha\alpha^{\prime}}^{i}\hat{L}_{\beta\beta^{\prime}}^{j}.\label{two-site}
\end{equation}

The Bose-Hubbard Hamiltonian \citep{gersch1963quantum,fisher1989boson}
\begin{equation}
\hat{H}=-J\sum_{\left\langle i,j\right\rangle }\hat{a}_{i}^{\dagger}\hat{a}_{j}+\frac{U}{2}\sum_{i}\hat{n}_{i}\left(\hat{n}_{i}-1\right)-\mu\sum_{i}\hat{n}_{i}\label{H}
\end{equation}
written in terms of creation ($\hat{a}_{i}^{\dagger}$), annihilation
($\hat{a}_{i}$) and number operators ($\hat{n}_{i}\equiv\hat{a}_{i}^{\dagger}\hat{a}_{i}$)
can be rewritten in a SBO form by means of Eqs. (\ref{single-site})
and (\ref{two-site}). The hopping term requires two-site representation,
while the rest is a sum of on-site terms:
\begin{equation}
\hat{H}=\sum_{i\alpha}E_{\alpha}\hat{L}_{\alpha\alpha}^{i}-J\sum_{\left\langle i,j\right\rangle }\sum_{\alpha\alpha^{\prime}\beta\beta^{\prime}}T_{\alpha\alpha^{\prime}\beta\beta^{\prime}}\hat{L}_{\alpha\alpha^{\prime}}^{i}\hat{L}_{\beta\beta^{\prime}}^{j}.\label{SBO H}
\end{equation}
Greek letter indices label Fock states just by the number of bosons.
The on-site contribution is given by
\begin{equation}
E_{\alpha}=\frac{U}{2}\alpha\left(\alpha-1\right)-\mu\alpha,\label{E}
\end{equation}
while the hopping matrix is
\begin{align}
T_{\alpha\alpha^{\prime}\beta\beta^{\prime}} & =\left\langle \alpha\right|\hat{a}^{\dag}\left|\alpha^{\prime}\right\rangle \left\langle \beta\right|\hat{a}\left|\beta^{\prime}\right\rangle \nonumber \\
 & =\sqrt{\alpha\beta^{\prime}}\delta_{\alpha,\alpha^{\prime}+1}\delta_{\beta+1,\beta^{\prime}}.\label{T matrix}
\end{align}

Since the Mott-insulator exhibits the $U\left(1\right)$ symmetry,
it is easier to approach the critical line from this side of the phase
diagram. The symmetry enforces various expectation values to vanish.
Invoking the symmetry transformation rules: $\hat{a}\rightarrow e^{-\mathrm{i}\varphi}\hat{a}$,
$\hat{a}^{\dag}\rightarrow e^{\mathrm{i}\varphi}\hat{a}^{\dag}$,
$\left|\alpha\right\rangle \rightarrow e^{\mathrm{i}\alpha\varphi}\left|\alpha\right\rangle $,
leads to $\hat{L}_{\alpha\beta}^{i}\rightarrow e^{\mathrm{i}\left(\alpha-\beta\right)\varphi}\hat{L}_{\alpha\beta}^{i}$.
Therefore, the unbroken $U\left(1\right)$ symmetry implies $\left\langle \hat{L}_{\alpha\beta}^{i}\right\rangle =0$
for $\alpha\neq\beta$. The remaining non-vanishing expectation values
can be denoted by
\begin{equation}
D_{\alpha}\equiv\left\langle \hat{L}_{\alpha\alpha}^{i}\right\rangle \label{D}
\end{equation}
and interpreted as boson occupancies. In other words, $D_{\alpha}$
is the probability of finding exactly $\alpha$ bosons on a given
site. Equation (\ref{D}) assumes a homogenous system with unbroken
translational symmetry.

Following Zubarev \citep{Zubarev1974} and Nolting \citep{Nolting2009},
we introduce retarded and advanced Green's functions for arbitrary
operators $\hat{A}$ and $\hat{B}$ as follows:
\begin{equation}
G_{AB}^{\pm}\left(t\right)=\mp\mathrm{i}\theta\left(\pm t\right)\left\langle \left[\hat{A}\left(t\right),\hat{B}\left(0\right)\right]\right\rangle ,\label{G}
\end{equation}
where $+$ corresponds to the retarded version, $-$ corresponds to
the advanced one, $\theta\left(t\right)$ is the Heaviside step function
and $\hat{A}\left(t\right),\hat{B}\left(0\right)$ refer to the Heisenberg
evolution of these operators. The equation of motion for such a Green's
function is obtained by differentiating Eq. (\ref{G}) with respect
to time. After going to the frequency representation via Fourier transform,
equation of motion becomes
\begin{equation}
\omega G_{AB}^{\pm}\left(\omega\right)=\left\langle \left[\hat{A}\left(0\right),\hat{B}\left(0\right)\right]\right\rangle -G_{\left[H,A\right],B}^{\pm}\left(\omega\right),\label{EoM}
\end{equation}
where the last Green's function is based on operators $\left[\hat{H},\hat{A}\right]$
and $\hat{B}$. Another formula that will be useful later is the spectral
theorem \citep{sajna2015ground,Zubarev1974,Nolting2009}, which allows
to calculate expectation values of operator products from the Green's
function:
\begin{align}
\left\langle \hat{A}\left(t\right)\hat{B}\left(0\right)\right\rangle  & =\int_{-\infty}^{\infty}\frac{\mathrm{d}\omega}{2\pi\mathrm{i}}e^{\mathrm{i}\omega t}f\left(\omega\right)\nonumber \\
 & \times\left[G_{BA}\left(\omega-\mathrm{i}0^{-}\right)-G_{BA}\left(\omega+\mathrm{i}0^{+}\right)\right].\label{spectral-2}
\end{align}
$f\left(\omega\right)=\left(e^{\omega/T}-1\right)^{-1}$ is the Bose-Einstein
distribution (Boltzmann constant $k_{B}=1$ and $T$ stands for temperature).
$G_{BA}$ without a $\pm$ superscript is the joined Green's function
\citep{Zubarev1974}, which equals the retarded one at the upper complex
half-plane and the advanced one at the lower complex half-plane. Formula
(\ref{spectral-2}) is more conveniently evaluated using Cauchy integral
theorem and residues of the Green's function:
\begin{equation}
\left\langle \hat{A}\left(t\right)\hat{B}\left(0\right)\right\rangle =\sum_{n}\mathrm{Res}_{\omega=\omega_{n}}\left\{ G_{BA}\left(\omega\right)\right\} f\left(\omega_{n}\right)e^{\mathrm{i}\omega_{n}t}.\label{spectral-1}
\end{equation}

\section{Self-consistency equations\label{sec:Self-consistency-equations}}

The aim of this section is to write Eq. (\ref{EoM}) for $\hat{A}=\hat{a}_{i}$,
$\hat{B}=\hat{a}_{j}^{\dag}$, rewrite it in terms of SBO and use
Random Phase Approximation (RPA) to decouple $G_{\left[H,A\right],B}$
in terms of $G_{AB}$, to finally obtain a solvable equation for $G_{AB}$
alone. The considered Green's function
\begin{equation}
G^{ij\pm}\left(t\right)=\mp\mathrm{i}\theta\left(\pm t\right)\left\langle \left[\hat{a}_{i}\left(t\right),\hat{a}_{j}^{\dag}\left(0\right)\right]\right\rangle \label{mainGreen}
\end{equation}
is expressible as a linear combination of SBO Green's functions:
\begin{equation}
G^{ij\pm}\left(t\right)=\sum_{\alpha\alpha^{\prime}\beta\beta^{\prime}}\left\langle \alpha\right|\hat{a}\left|\alpha^{\prime}\right\rangle \left\langle \beta\right|\hat{a}^{\dagger}\left|\beta^{\prime}\right\rangle G_{\alpha\alpha^{\prime}\beta\beta^{\prime}}^{ij\pm}\left(t\right),\label{mainGreenRecipe}
\end{equation}
with
\begin{equation}
G_{\alpha\alpha^{\prime}\beta\beta^{\prime}}^{ij\pm}\left(t\right)=\mp\mathrm{i}\theta\left(\pm t\right)\left\langle \left[\hat{L}_{\alpha\alpha^{\prime}}^{i}\left(t\right),\hat{L}_{\beta\beta^{\prime}}^{j}\left(0\right)\right]\right\rangle .
\end{equation}

The commutator of a SBO with the Hamiltonian is
\begin{align}
 & \left[\hat{L}_{\alpha\alpha^{\prime}}^{i},\hat{H}\right]=\nonumber \\
 & -\left(E_{\alpha}-E_{\alpha^{\prime}}\right)\hat{L}_{\alpha\alpha^{\prime}}^{i}-J\sum_{r\text{n.t.}i}\sum_{\bm{\alpha}_{1}\bm{\beta}_{1}}\left(T_{\bm{\beta}_{1}\bm{\alpha}_{1}}+T_{\bm{\alpha}_{1}\bm{\beta}_{1}}\right)\nonumber \\
 & \times\left(\hat{L}_{\alpha\beta_{1}^{\prime}}^{i}\delta_{\alpha^{\prime}\beta_{1}}-\hat{L}_{\beta_{1}\alpha^{\prime}}^{i}\delta_{\beta_{1}^{\prime}\alpha}\right)\hat{L}_{\bm{\alpha}_{1}}^{r},\label{=00005B.,.=00005D}
\end{align}
where $\sum_{r\text{n.t.}i}$ denotes summation over all nearest neighbors
$r$ of site $i$ and bold greek indices denote double indices, e.
g. $\bm{\alpha}=\left(\alpha,\alpha^{\prime}\right)$.

Writing Eq. (\ref{EoM}) for $\hat{A}=\hat{L}_{\alpha\alpha^{\prime}}^{i}$
and $\hat{B}=\hat{L}_{\beta\beta^{\prime}}^{j}$ gives
\begin{align}
\omega G_{\bm{\alpha\beta}}^{ij}\left(\omega\right) & =\left\langle \left[\hat{L}_{\alpha\alpha^{\prime}}^{i}\left(0\right),\hat{L}_{\beta\beta^{\prime}}^{i}\left(0\right)\right]\right\rangle \delta_{ij}\nonumber \\
 & -G_{\left[H,L_{\alpha\alpha^{\prime}}^{i}\right],L_{\beta\beta^{\prime}}^{j}}\left(\omega\right).\label{EoM_step}
\end{align}
$G_{AB}\left(\omega\right)$ is bilinear with respect to $A$ and
$B$, so $G_{\left[H,L_{\alpha\alpha^{\prime}}^{i}\right],L_{\beta\beta^{\prime}}^{j}}\left(\omega\right)$
can be expanded using Eq. (\ref{=00005B.,.=00005D}). To express the
result compactly, a higher order Green's function can be introduced:
\begin{equation}
G_{\bm{\alpha}\bm{\gamma}\bm{\beta}}^{ij\pm}\left(t\right)\equiv\mp\mathrm{i}\theta\left(\pm t\right)\sum_{r\text{n.t.}i}\left\langle \left[\hat{L}_{\bm{\alpha}}^{i}\left(t\right)\hat{L}_{\bm{\gamma}}^{r}\left(t\right),\hat{L}_{\bm{\beta}}^{j}\left(0\right)\right]\right\rangle .\label{3G}
\end{equation}
Let $G_{\bm{\alpha}\bm{\gamma}\bm{\beta}}^{ij}\left(\omega\right)$
be its Fourier transform in time. Then, Eq. (\ref{EoM_step}) becomes
\begin{align}
 & \left(\omega+E_{\alpha}-E_{\alpha^{\prime}}\right)G_{\bm{\alpha\beta}}^{ij}\left(\omega\right)\nonumber \\
 & +J\sum_{\bm{\mu}\bm{\alpha}_{1}\bm{\beta}_{1}}\left(T_{\bm{\beta}_{1}\bm{\alpha}_{1}}+T_{\bm{\alpha}_{1}\bm{\beta}_{1}}\right)\nonumber \\
 & \times\left(\delta_{\mu\alpha}\delta_{\mu^{\prime}\beta_{1}^{\prime}}\delta_{\alpha^{\prime}\beta_{1}}-\delta_{\mu\beta_{1}}\delta_{\mu^{\prime}\alpha^{\prime}}\delta_{\beta_{1}^{\prime}\alpha}\right)G_{\bm{\mu}\bm{\alpha}_{1}\bm{\beta}}^{ij}\left(\omega\right)\nonumber \\
 & =\left(\left\langle \hat{L}_{\alpha\beta^{\prime}}\right\rangle \delta_{\alpha^{\prime}\beta}-\left\langle \hat{L}_{\beta\alpha^{\prime}}\right\rangle \delta_{\beta^{\prime}\alpha}\right)\delta_{ij}.\label{EoM_step2}
\end{align}

Now, a certain approximation is needed to decouple $G_{\bm{\mu}\bm{\alpha}_{1}\bm{\beta}}^{ij}$
into $G_{\bm{\alpha\beta}}^{ij}$, so that Eq. (\ref{EoM_step2})
can be solved for $G_{\bm{\alpha\beta}}^{ij}$. Here Random Phase
Approximation (RPA) comes into play. For arbitrary operators $\hat{A}$,
$\hat{B}$ and $\hat{C}$ it is given by the following rule:\citep{sajna2015ground}
\begin{equation}
\left\langle \left[\hat{A}\hat{B},\hat{C}\right]\right\rangle \overset{\mathrm{RPA}}{\approx}\left\langle \hat{A}\right\rangle \left\langle \left[\hat{B},\hat{C}\right]\right\rangle +\left\langle \left[\hat{A},\hat{C}\right]\right\rangle \left\langle \hat{B}\right\rangle .\label{RPA}
\end{equation}
Its discussion is reserved for Appendix \ref{sec:Conceptual-problems-with}.
Keeping in mind that $G_{\bm{\mu}\bm{\alpha}_{1}\bm{\beta}}^{ij}$
entering Eq. (\ref{EoM_step2}) always has nondiagonal $\bm{\alpha}_{1}$
and $\bm{\beta}$ (i. e. $\alpha_{1}\neq\alpha_{1}^{\prime}$ and
$\beta\neq\beta^{\prime}$), it is decoupled to
\begin{equation}
G_{\bm{\mu}\bm{\alpha}_{1}\bm{\beta}}^{ij}\left(\omega\right)\overset{\mathrm{RPA}}{\cong}\left\langle \hat{L}_{\bm{\mu}}\right\rangle \sum_{r\text{n.t.}i}G_{\bm{\alpha}_{1}\bm{\beta}}^{rj}\left(\omega\right).\label{decoupling}
\end{equation}
With Eq. (\ref{decoupling}), Eq. (\ref{EoM_step2}) becomes translationally
invariant, but not on-site (i. e. Green's functions $G_{\bm{\alpha\beta}}^{ij}$
and $G_{\bm{\alpha}_{1}\bm{\beta}}^{rj}$ both enter a single equation).
Performing a discrete Fourier transform in space and using the fact
that only diagonal elements of $\left\langle \hat{L}_{\bm{\alpha}}\right\rangle $
are nonzero (Eq. (\ref{D})) leads to
\begin{align}
 & \left(\omega+E_{\alpha}-E_{\alpha^{\prime}}\right)G_{\bm{\alpha\beta}}\left(\omega,k\right)\nonumber \\
 & +J\sum_{\bm{\alpha}_{1}}\left(T_{\alpha^{\prime}\alpha\bm{\alpha}_{1}}+T_{\bm{\alpha}_{1}\alpha^{\prime}\alpha}\right)\left(D_{\alpha}-D_{\alpha^{\prime}}\right)\epsilon\left(k\right)G_{\bm{\alpha}_{1}\bm{\beta}}\left(\omega,k\right)\nonumber \\
 & =\left(D_{\alpha}-D_{\alpha^{\prime}}\right)\delta_{\alpha\beta^{\prime}}\delta_{\alpha^{\prime}\beta}.\label{EoM_step4}
\end{align}
Since we are interested in the Green's function from Eq. (\ref{mainGreen})
given by Eq. (\ref{mainGreenRecipe}), we can restrict to $\bm{\alpha}\in\uparrow$
(which is a shorthand for $\alpha^{\prime}=\alpha+1$) and $\bm{\beta}\in\downarrow$
(which means $\beta^{\prime}=\beta-1$). Then $T_{\bm{\alpha}_{1}\alpha^{\prime}\alpha}=0$,
so using Eq. (\ref{T matrix}) simplifies Eq. (\ref{EoM_step4}) to
\begin{equation}
\sum_{\bm{\alpha}_{1}\in\uparrow}M_{\alpha\alpha_{1}}G_{\bm{\alpha}_{1}\bm{\beta}}\left(\omega,k\right)=\left(D_{\alpha}-D_{\alpha^{\prime}}\right)\delta_{\alpha\beta^{\prime}}\delta_{\alpha^{\prime}\beta},\label{EoM_step5}
\end{equation}
where
\begin{align}
M_{\alpha\alpha_{1}} & =\left(\omega+E_{\alpha}-E_{\alpha^{\prime}}\right)\delta_{\alpha\alpha_{1}}\nonumber \\
 & +J\sqrt{\left(\alpha+1\right)\left(\alpha_{1}+1\right)}\left(D_{\alpha}-D_{\alpha^{\prime}}\right)\epsilon\left(k\right).\label{M}
\end{align}

Inverting matrix $M$ allows to find the Green's function:
\begin{align}
G_{\bm{\alpha}_{1}\bm{\beta}}\left(\omega,k\right) & =\sum_{\bm{\alpha}\in\uparrow}\left[M^{-1}\left(\omega,k\right)\right]_{\alpha_{1}\alpha}\left(D_{\alpha}-D_{\alpha^{\prime}}\right)\delta_{\alpha\beta^{\prime}}\delta_{\alpha^{\prime}\beta}\nonumber \\
 & =\left[M^{-1}\left(\omega,k\right)\right]_{\alpha_{1}\beta^{\prime}}\left(D_{\beta^{\prime}}-D_{\beta}\right).\label{Gsol}
\end{align}
In principle, matrix $M$ has infinite size. It can be truncated by
introducing the lower ($\mathcal{M}$) and upper ($\mathcal{N}$)
cut-off for the number of bosons on each site. Then both indices $\alpha$
and $\alpha_{1}$ in $M_{\alpha\alpha_{1}}$ take values $\mathcal{M},\mathcal{M}+1,\dots,\mathcal{N}-1$.

Formula (\ref{Gsol}) yields a specific Green's function for given
occupancies $D_{\alpha}$. However, they are not a priori known. The
spectral theorem from Eq. (\ref{spectral-1}) enables to calculate
$D_{\alpha}$ back from the Green's function, allowing to set up self-consistency
equations. While there are actually many ways to do it (discussion
is left for Appendix \ref{sec:Conceptual-problems-with}), it is checked
in \citep{sajna2015ground} that the best way consists in setting
\begin{align}
 & \sqrt{\beta}D_{\beta}=\left\langle \hat{L}_{\beta,\beta-1}\hat{a}\right\rangle =\left\langle \hat{L}_{\beta,\beta-1}\left(\sum_{\alpha}\sqrt{\alpha}\hat{L}_{\alpha-1,\alpha}\right)\right\rangle \nonumber \\
 & =\sum_{\alpha}\frac{\sqrt{\alpha}}{N}\sum_{k}\sum_{n}f\left(\omega_{n}\right)\mathrm{Res}_{\omega=\omega_{n}}\left\{ G_{\alpha-1,\alpha,\beta,\beta-1}\left(\omega,k\right)\right\} ,\label{sc-eq}
\end{align}
where $N$ is the total number of sites. Inserting Eq. (\ref{Gsol})
to Eq. (\ref{sc-eq}) gives
\begin{align}
 & \sqrt{\beta}D_{\beta}=\left(D_{\beta-1}-D_{\beta}\right)\sum_{\alpha}\frac{\sqrt{\alpha}}{N}\nonumber \\
 & \times\sum_{kn}f\left(\omega_{n}\right)\mathrm{Res}_{\omega=\omega_{n}}\left\{ \left[M^{-1}\left(\omega,k\right)\right]_{\alpha-1,\beta-1}\right\} ,\label{sc-eq2}
\end{align}
where $\alpha,\beta\in\left\{ \mathcal{M}+1,\dots,\mathcal{N}\right\} $.
Thus, there is $\mathcal{N}-\mathcal{M}$ equations for $\mathcal{N}-\mathcal{M}+1$
unknowns ($D_{\mathcal{M}},\dots,D_{\mathcal{N}}$). The missing equation
is naturally provided by the normalization condition
\begin{equation}
\sum_{\alpha=\mathcal{M}}^{\mathcal{N}}D_{\alpha}=1.\label{norm}
\end{equation}

In the thermodynamic limit, as the number of sites $N$ tends to infinity,
$N^{-1}\sum_{k}$ turns into an integral over the $k$-space. Since
$M^{-1}\left(\omega,k\right)$ depends on $k$ only through $\epsilon\left(k\right)$,
it is convenient to exchange integration over $k$ for integration
over $\epsilon$. This is done with the help of density of states
$\rho\left(x\right)$. It is defined so that $\epsilon=2x$ and $x\in\left[-d,d\right]$
($d=3$ is the dimensionality).

Introducing matrix $X$ via
\begin{equation}
X_{\alpha\beta}\left(\epsilon,D\right)\equiv\sum_{n}f\left(\omega_{n}\right)\mathrm{Res}_{\omega=\omega_{n}}\left\{ \left[M^{-1}\left(\omega,\epsilon,D\right)\right]_{\alpha\beta}\right\} \label{X}
\end{equation}
and using Eq. (\ref{sc-eq2}), the final equations become
\begin{align}
 & \sqrt{\beta}D_{\beta}=\left(D_{\beta-1}-D_{\beta}\right)\sum_{\alpha}\sqrt{\alpha}\nonumber \\
 & \times\int_{-d}^{d}\mathrm{d}x\,\rho_{s}\left(x\right)X_{\alpha-1,\beta-1}\left(2x,D\right),\label{final}
\end{align}
for $\alpha,\beta\in\left\{ \mathcal{M}+1,\dots,\mathcal{N}\right\} $,
augmented by Eq. (\ref{norm}).

\section{Results for single-component Bose-Hubbard model\label{sec:Results-for-single-component}}

\begin{figure}[H]
\centering{}\includegraphics[scale=0.42]{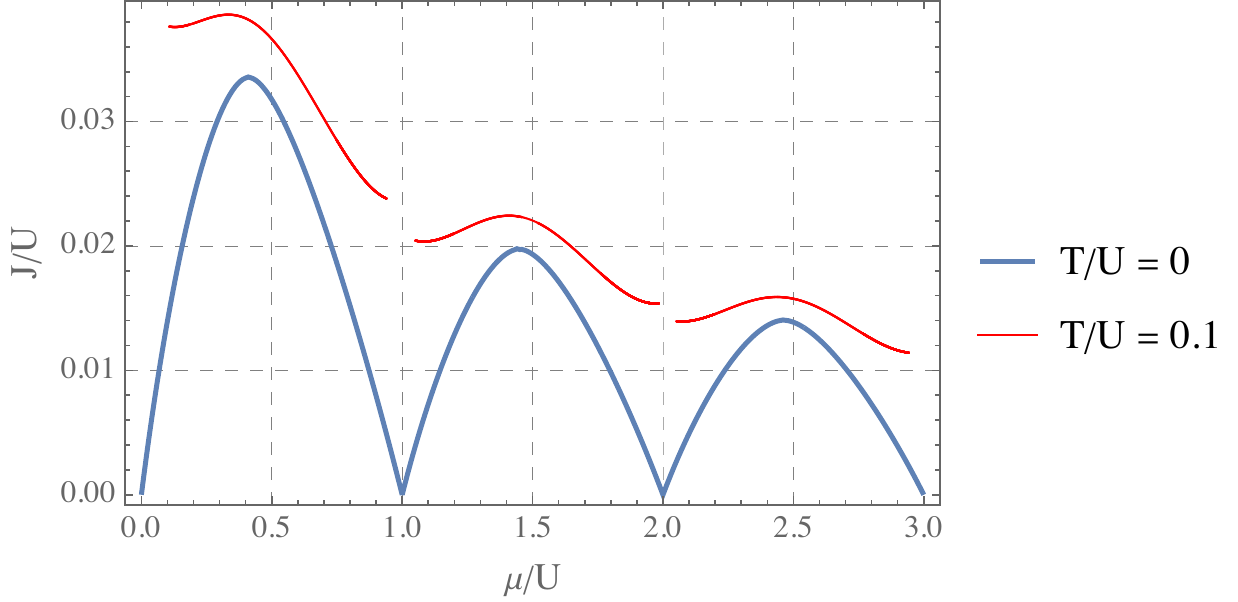}\caption{Three lobes in zero and moderate temperature in the TSA.\label{fig:Three-lobes-in}}
\end{figure}

The previous usage of the SBO method already gave high consistency
with Monte-Carlo results at zero temperature.\citep{sajna2015ground}
However, plotting the lobes at moderate temperatures (i. e. such that
the lobe structure is still pronounced) reveals unphysical discontinuities
(Fig. \ref{fig:Three-lobes-in}), which are not predicted by different
methods \citep{martin2025finite}. The reason behind this spurious
effect is the three-states approximation (TSA). If $n=\left\lceil \mu/U\right\rceil $,
then the Fock states of $n-1$, $n$ and $n+1$ bosons are used. Thus
transitioning between lobes is associated with a sudden change of
the basis and thus occupancies. These jumps in the critical hopping
parameter are very large when compared to the discrepancies from Monte-Carlo
results.

\begin{figure}
\centering{}\includegraphics[scale=0.43]{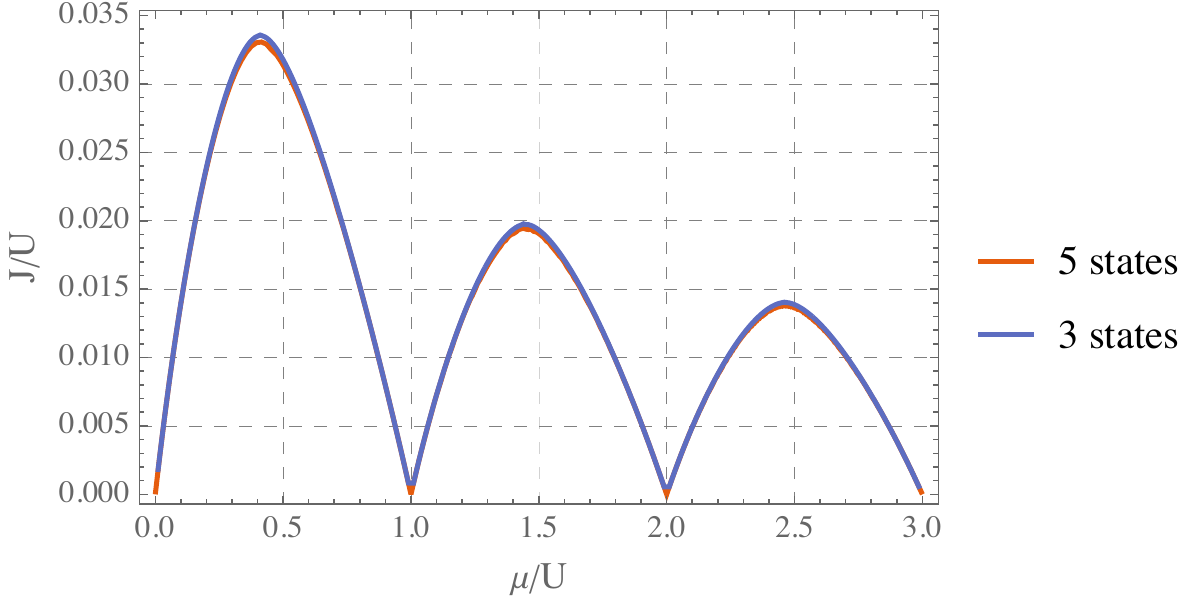}\caption{Comparison of the zero-temperature lobes for three- and five-states
approximation.\label{fig:Comparison-of-the}}
\end{figure}

Using the developed formalism with a systematically extendable local
Hilbert space, a large enough basis set can be chosen, so that there
is no need to change it while transitioning between adjacent lobes.
Even if such a change is necessary, only the peripheral basis elements
are added or removed, which are very little populated, so that discontinuities
are negligible. At zero temperature, using systematically extendable
local Hilbert-space description changes the results marginally (Fig.
\ref{fig:Comparison-of-the}). More precisely, $\mathcal{M}=0$ and
$\mathcal{N}=4$ was used for the first two lobes, while $\mathcal{M}=1$
and $\mathcal{N}=5$ for the third. Occupation of the maximal state
was on the order of $10^{-4}$, while $D_{1}$ reached maximally $1.2\times10^{-3}$
for the third lobe. This analysis ensures that no more states are
needed at $T=0$. Funnily, the correction brought by the inclusion
of more states is in the opposite direction than needed to match Monte-Carlo
results. Anyway, the discrepancy is still very small staying less
than $3\%$.

\begin{figure}
\centering{}\includegraphics[scale=0.43]{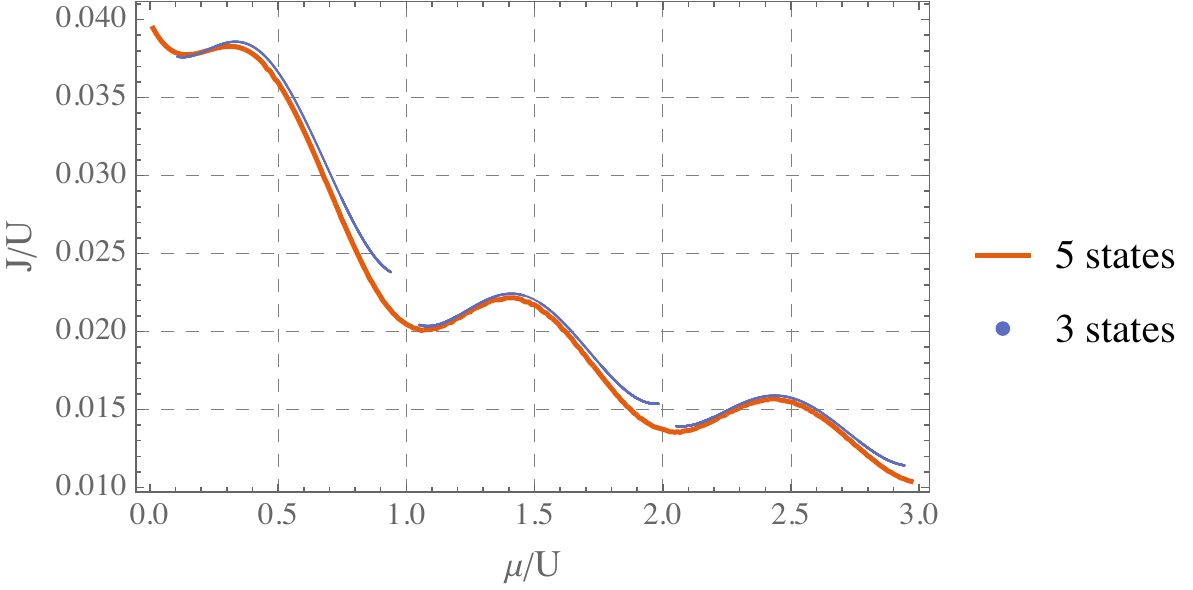}\caption{Comparison of the lobes at moderate temperature ($T/U=0.1$) for three
and five-states approximation.\label{fig:Comparison-of-the-T}}
\end{figure}

Finally, at nonzero temperature, the desired improvement is achieved
(Fig. \ref{fig:Comparison-of-the-T}). The critical line is smooth,
with a minor discontinuity at the transition from second to the third
lobe. Occupation of the maximal state (and the lowest for the third
lobe) was no greater than $3\times10^{-2}$, which again validates
the usage of only five states.

\begin{figure*}
\begin{centering}
\subfloat[Left side of the first lobe.\label{fig:Left-side-of}]{\centering{}\includegraphics[scale=0.24]{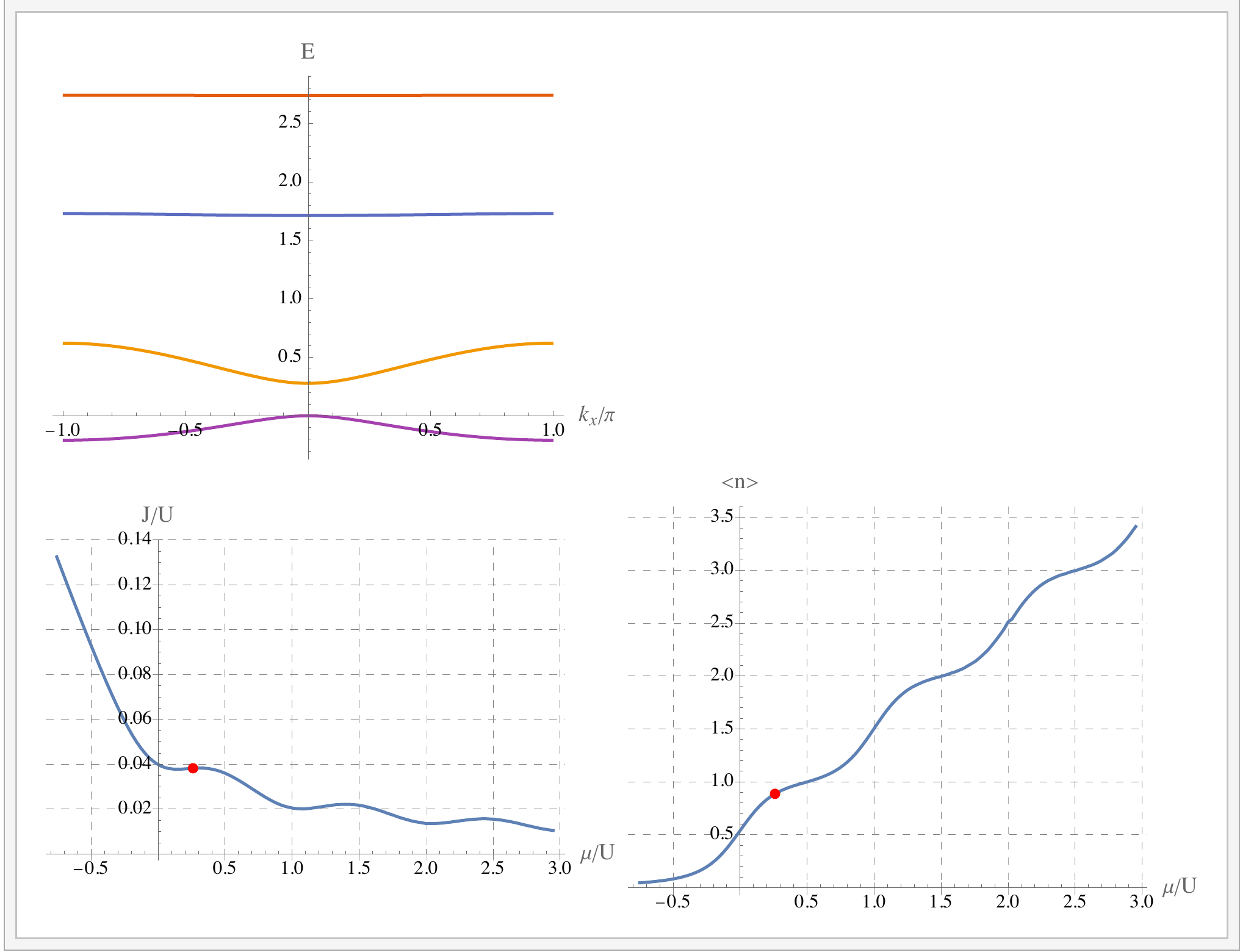}}\subfloat[Tip of the first lobe.\label{fig:Tip-of-the}]{\centering{}\includegraphics[scale=0.24]{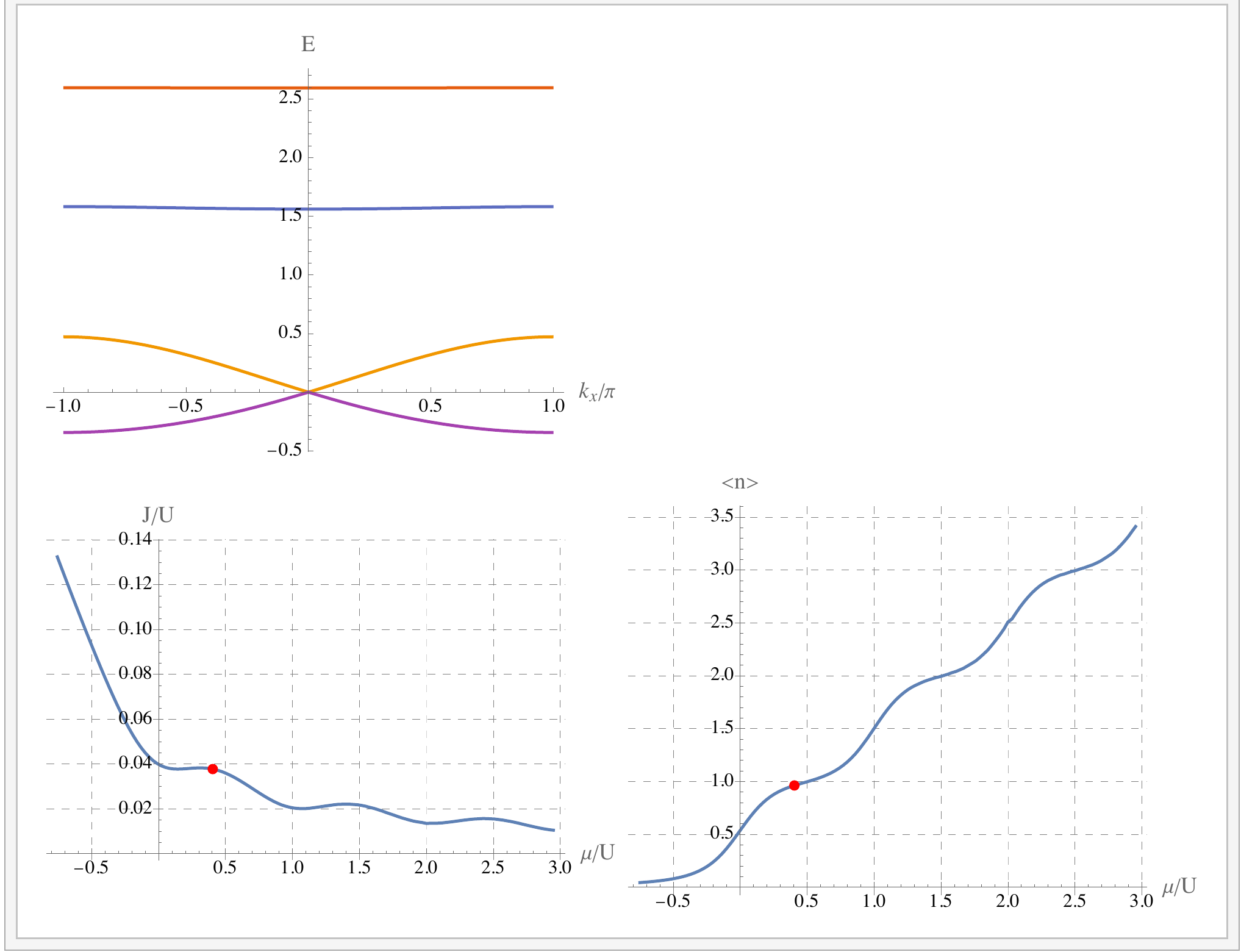}}\\
\par\end{centering}
\centering{}\subfloat[Right side of the first lobe.\label{fig:Right-side-of}]{\centering{}\includegraphics[scale=0.24]{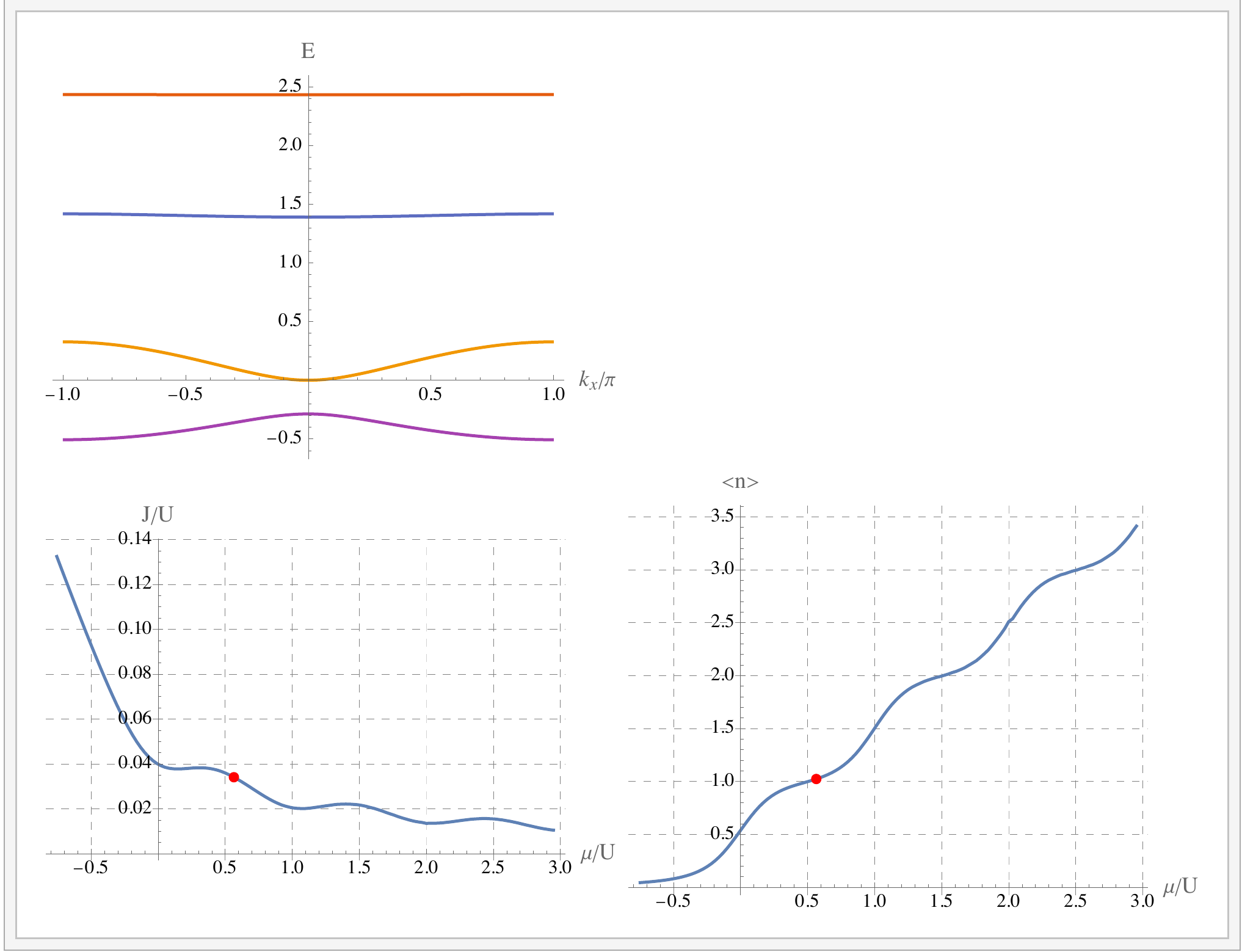}}\subfloat[Inter-lobe point.\label{fig:Inter-lobe-point.}]{\centering{}\includegraphics[scale=0.24]{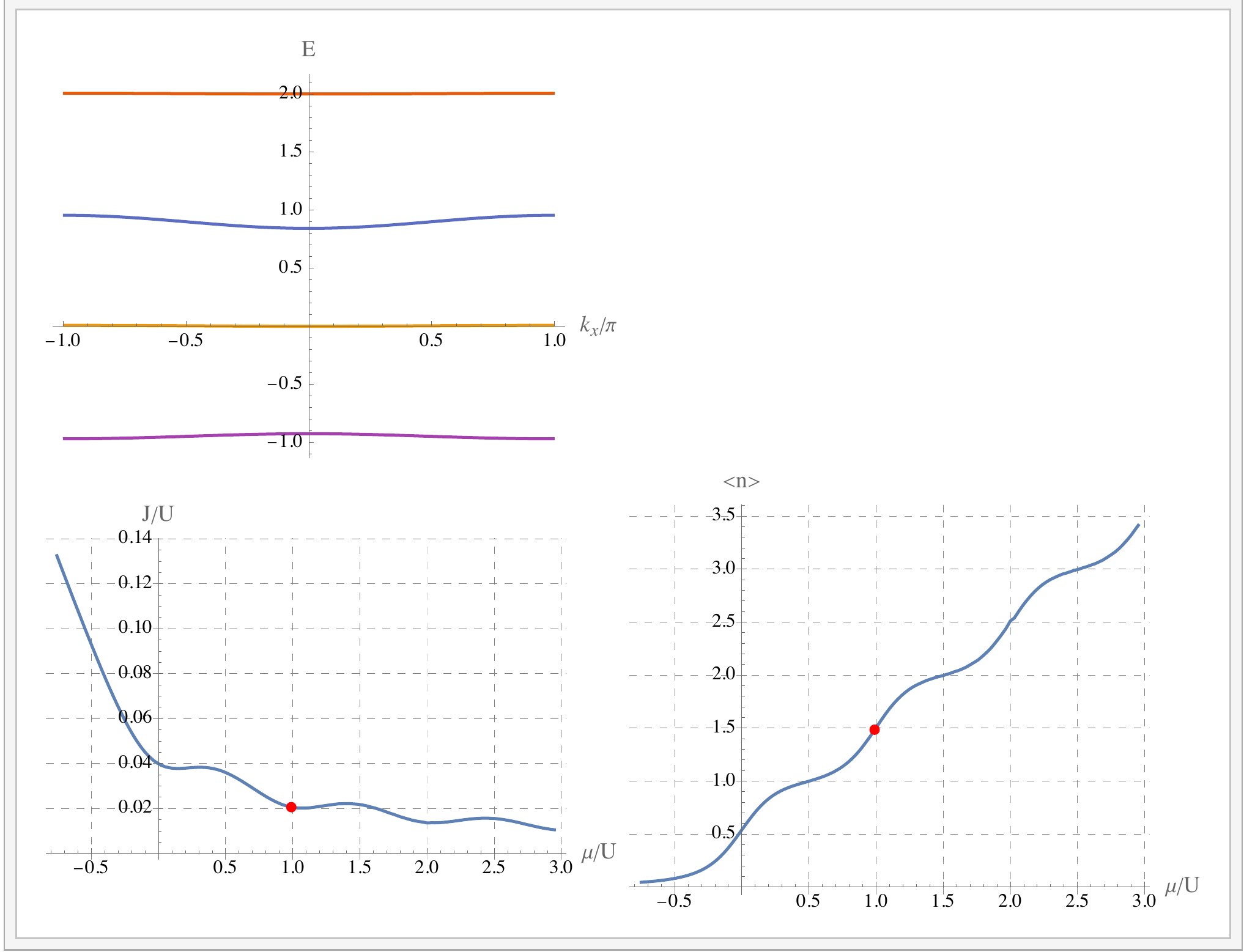}}\caption{Dispersion relation for various points on the critical line presented
together with the mean number of bosons per site as a function of
the chemical potential.\label{fig:Dispersion-relation-for}}
\end{figure*}

A very nice feature of the SBO formalism is its easy access to the
dispersion relation, while inclusion of many states introduces more
branches. Figure \ref{fig:Dispersion-relation-for} shows how the
dispersion relation changes at different points of the first lobe,
visualizing some of the properties invoked in Appendix \ref{sec:Critical-line-and}.
An additional advantage of the extended local basis is that the quasiparticle
spectrum evolves continuously as external parameters change. Instead
of introducing artificial changes associated with abrupt modifications
of the local Hilbert space, additional excitation branches naturally
emerge and evolve within a single self-consistent framework, providing
a physically consistent picture throughout the entire Mott region.

\begin{figure*}
\centering{}\includegraphics[scale=0.5]{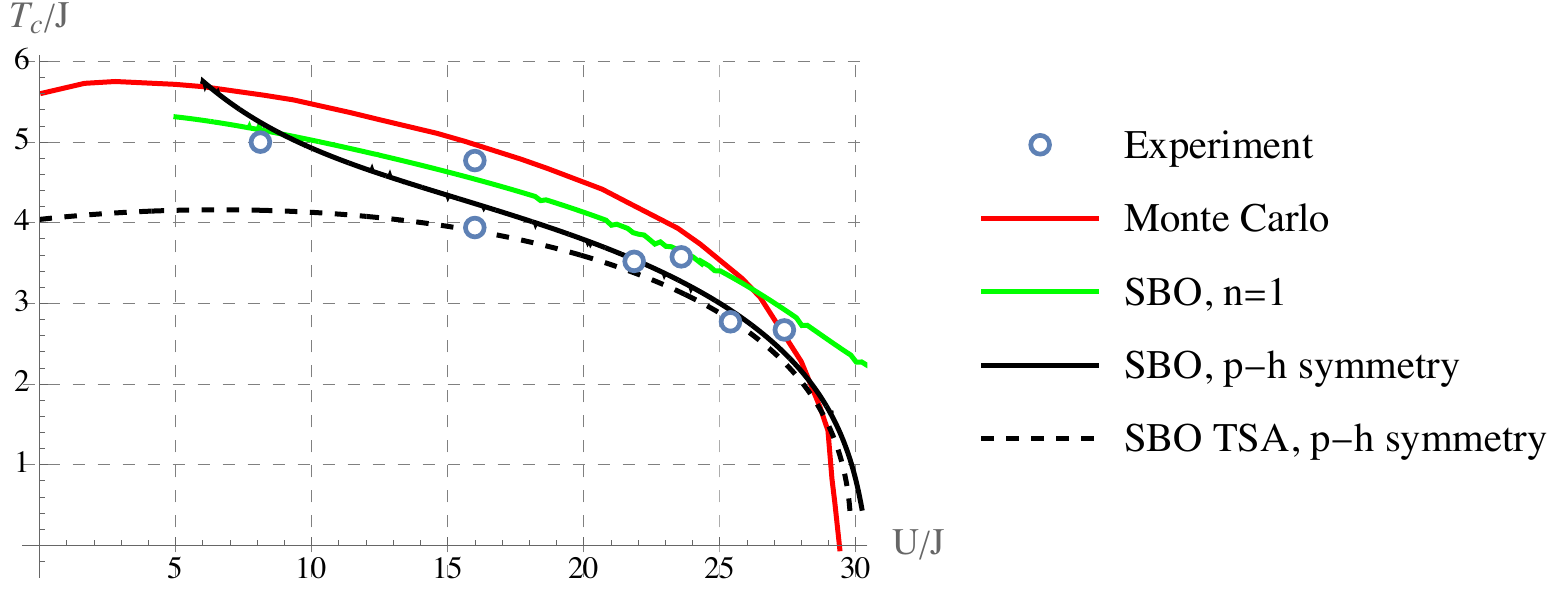}\caption{Critical temperature as a function of interaction strength for various
settings. The experimental data comes from \citep{trotzky2010suppression}.\label{fig:Critical-temperature-as-1}}
\end{figure*}

Figure \ref{fig:Critical-temperature-as-1} shows the critical temperature
$T_{c}/J$ as a function of the interaction strength $U/J$, normalized
by hopping. SBO method using the TSA underestimates the critical temperature,
when its values are higher. This is improved by including more states.
However, there is no single $T_{c}$ vs. $U$ curve, because it can
be plotted for different conditions. One choice is fixing the filling,
e. g. $n=1$, which was claimed in the experiment \citep{trotzky2010suppression}.
The other choice is fixing the chemical potential $\mu$, which was
not assumed for any presented plots.\textcolor{red}{{} }Finally, one
can take $\mu$ corresponding to the particle-hole symmetry. This
was assumed (while $\mu/U\in\left(0,1\right)$) in SBO results in
\inlinecite{sajna2015ground}. As can be seen from Fig. \ref{fig:Critical-temperature-as-1},
the curves obtained with controlled enlargement of the local Hilbert
space under different conditions have significantly different behavior.
They take similar values in the range of $U$ from the experiment,
but the black solid curve (plotted for particle-hole symmetry) suddenly
grows as $U/J$ is lowered below roughly $7$. The green line (done
for $n=1$) fails to cross the horizontal axis at the quantum critical
point. The reason behind the behavior of the black curve for small
$U$ comes from populating higher boson number states. Thus the dashed
curve is completely out of its applicability region already for $U/J\approx10$,
which corresponds to occupation of the highest state being $0.20$.
The black curve is plotted for its region of applicability, which
was determined so that the occupation of the highest state did not
exceed $0.015$. The reason behind high-$U$ behavior of the green
curve is simply very low compressibility of the Mott insulator (or
formally the normal gas phase). Changing $n$ slightly has an enormous
effect on $\mu$. Thus setting $n=1$ or $n=1.01$ would produce clearly
different curves in the low $T$ region. The SBO method correctly
predicts $n\approx1$ for any $\mu/U\in\left(0,1\right)$ at $T=0$,
but even a tiny deviation from $n=1$ would produce anomalies as one
fixes $n=1$ as a condition. Thus staying on the tip's lobe is much
more robust for low temperatures than the $n=1$ requirement. Moreover,
it is impossible to ensure $n=1$ in the experiment exactly. At low
$U$, the green curve ends when a transition to the zeroth lobe happens.

\begin{figure*}
\begin{centering}
\subfloat[]{\centering{}\includegraphics[scale=0.2]{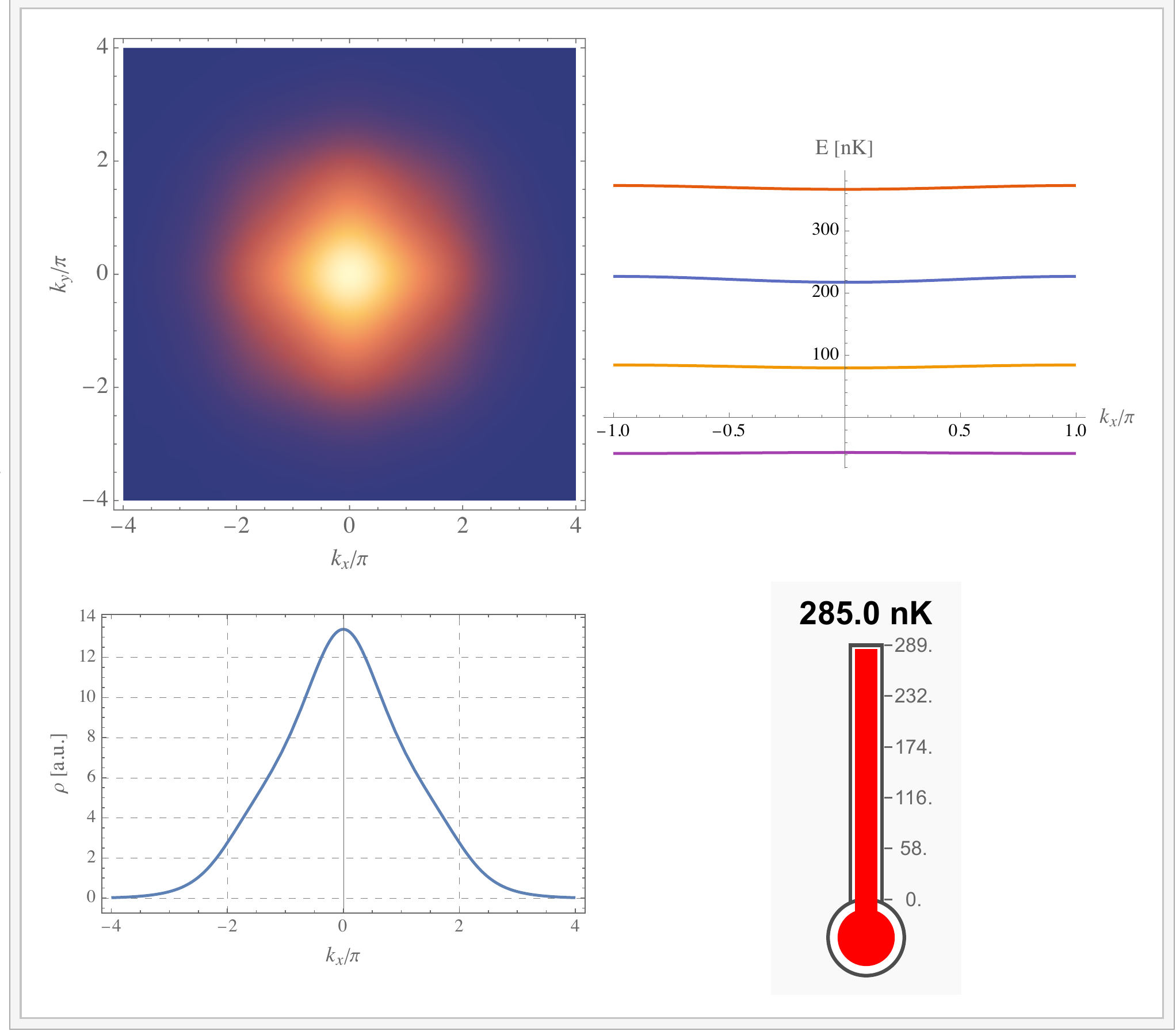}}\subfloat[]{\centering{}\includegraphics[scale=0.2]{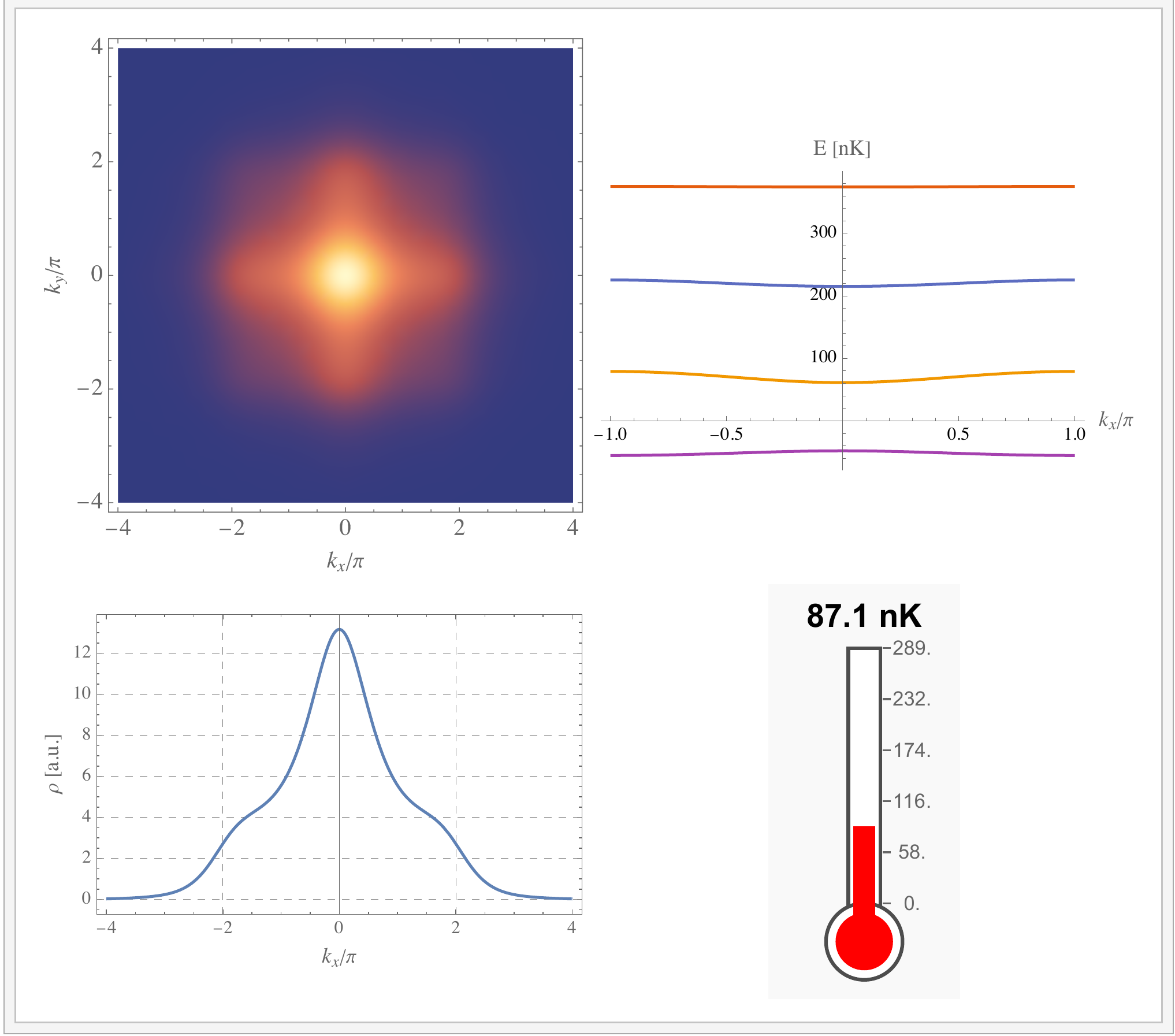}}\\
\par\end{centering}
\begin{centering}
\subfloat[]{\centering{}\includegraphics[scale=0.2]{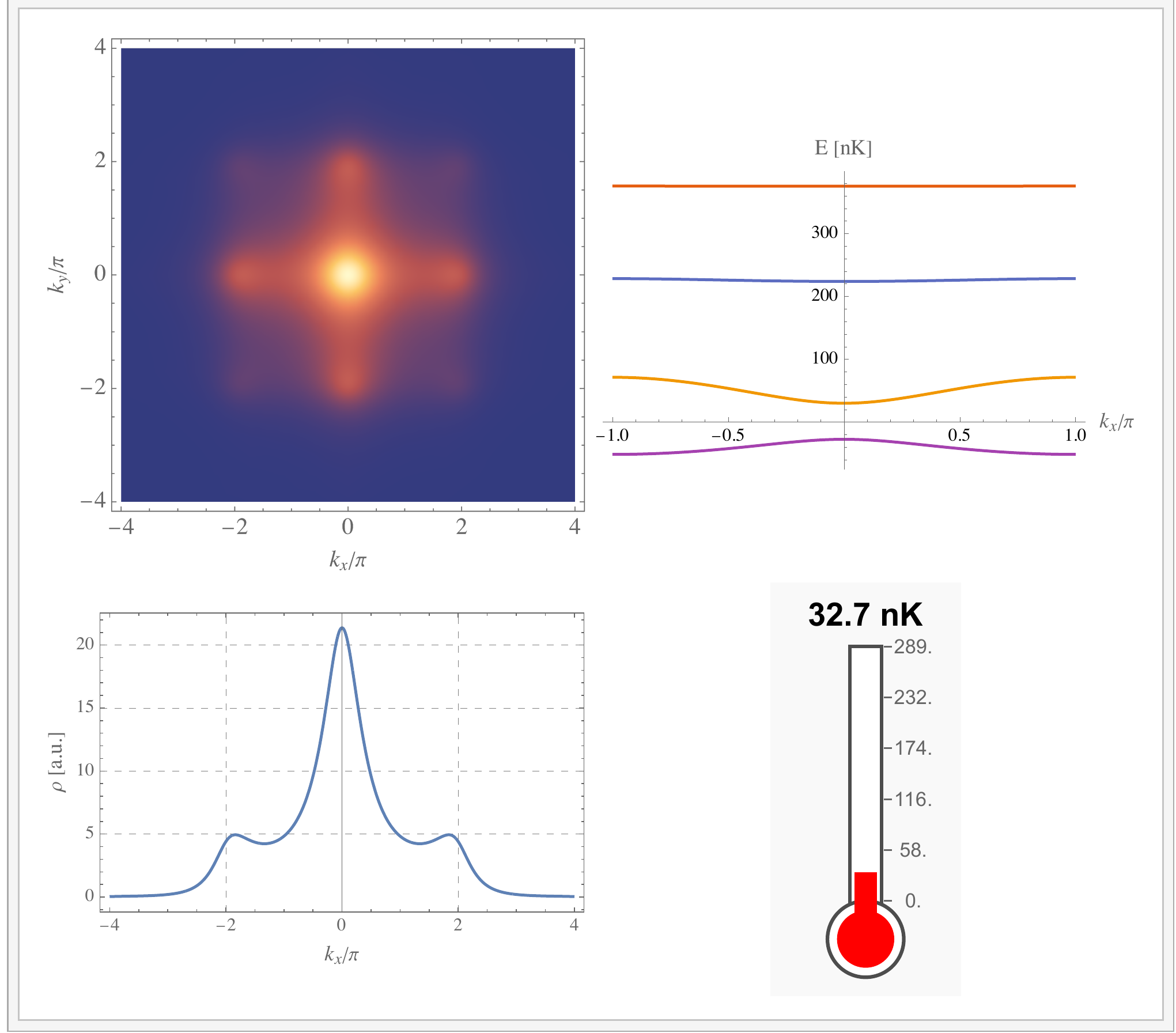}}\subfloat[]{\centering{}\includegraphics[scale=0.2]{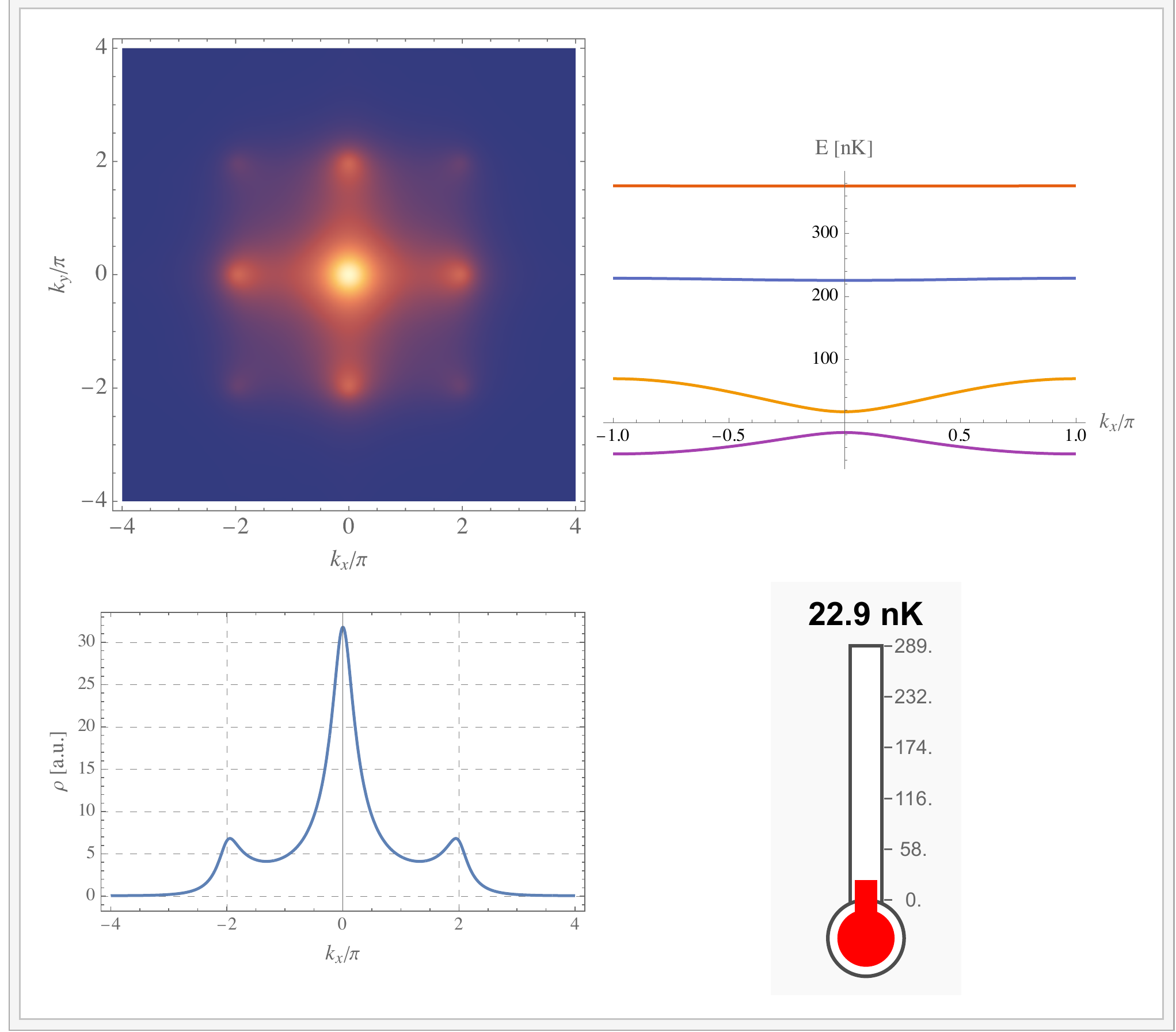}}\\
\par\end{centering}
\centering{}\subfloat[]{\centering{}\includegraphics[scale=0.2]{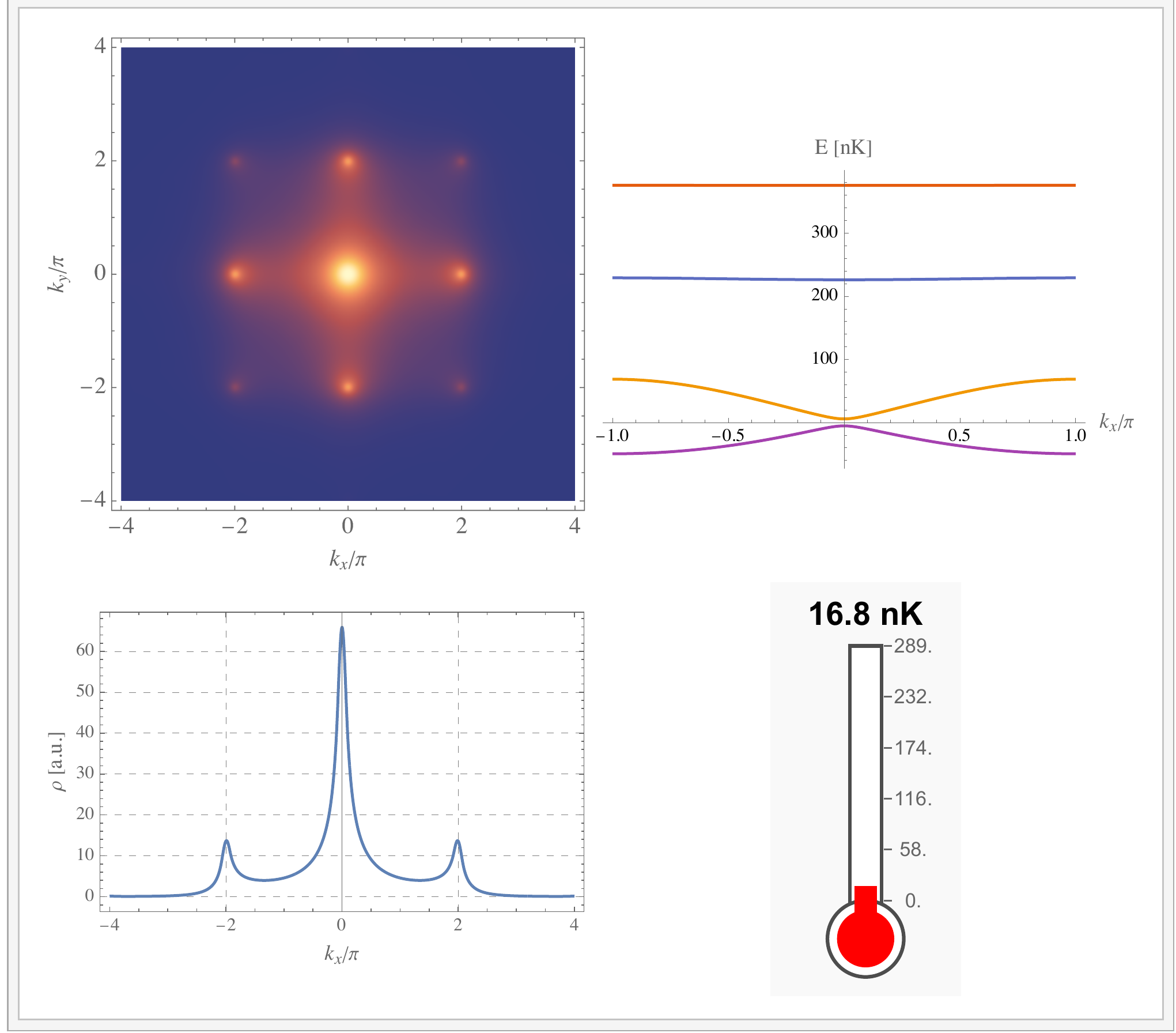}}\subfloat[]{\centering{}\includegraphics[scale=0.2]{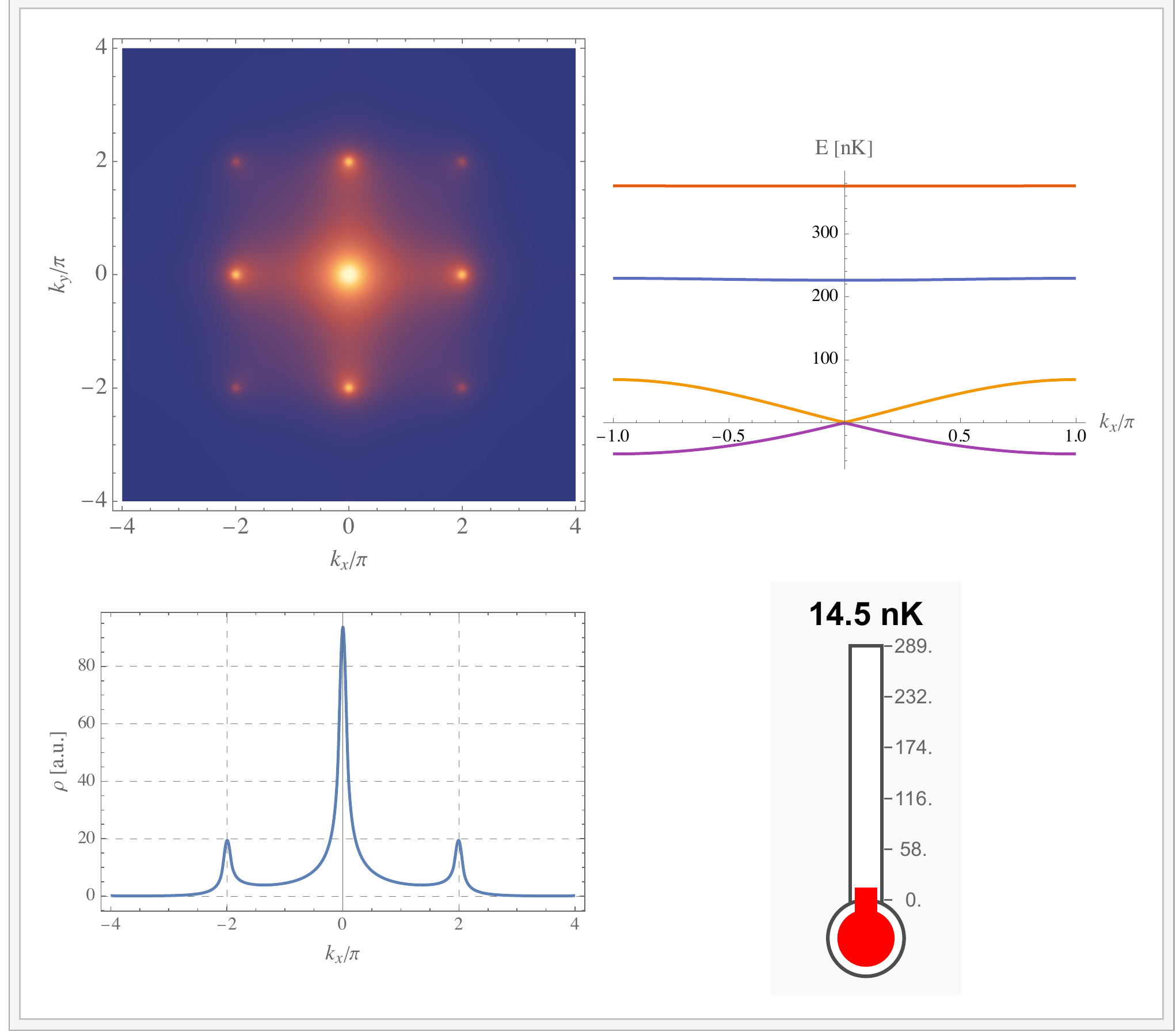}}\caption{Time of flight patterns as a function of temperature presented together
with the dispersion plots and reference thermometer. \label{fig:Time-of-flight}}
\end{figure*}

Figure \ref{fig:Time-of-flight} shows the generated time of flight
patterns on the basis of standard formulas\citep{hoffmann2009visibility,kashurnikov2002revealing}.
The lowest temperature $T/U=0.1$ corresponds to a critical point
with particle-hole symmetry having $\mu/U=0.40$ and $J/U=0.038$.
Higher temperatures (with unchanged $U$, $\mu$ and $J$) exhibit
less pronounced interference peaks and opened gap in the dispersion
relation. A $k_{y}=0$ slice is presented below the two-dimensional
absorption density profile. Also, a reference thermometer is shown,
which is scaled by taking a value of $U=2\pi\times480\,\mathrm{Hz}$
from \citep{preiss2016atomic}. This is done only to emphasize the
temperature scale in SI units. Also, value of $w_{0}$ was taken as
$0.2$, to qualitatively reproduce strength of satellite peaks. The
observed evolution should not be interpreted merely as thermal broadening
of interference peaks. Since the momentum distribution is reconstructed
from the underlying Green’s functions, the changes also reflect redistribution
of spectral weight associated with thermal occupation of higher local
states and corresponding modifications of the quasiparticle spectrum.

\begin{figure}
\centering{}\includegraphics[scale=0.42]{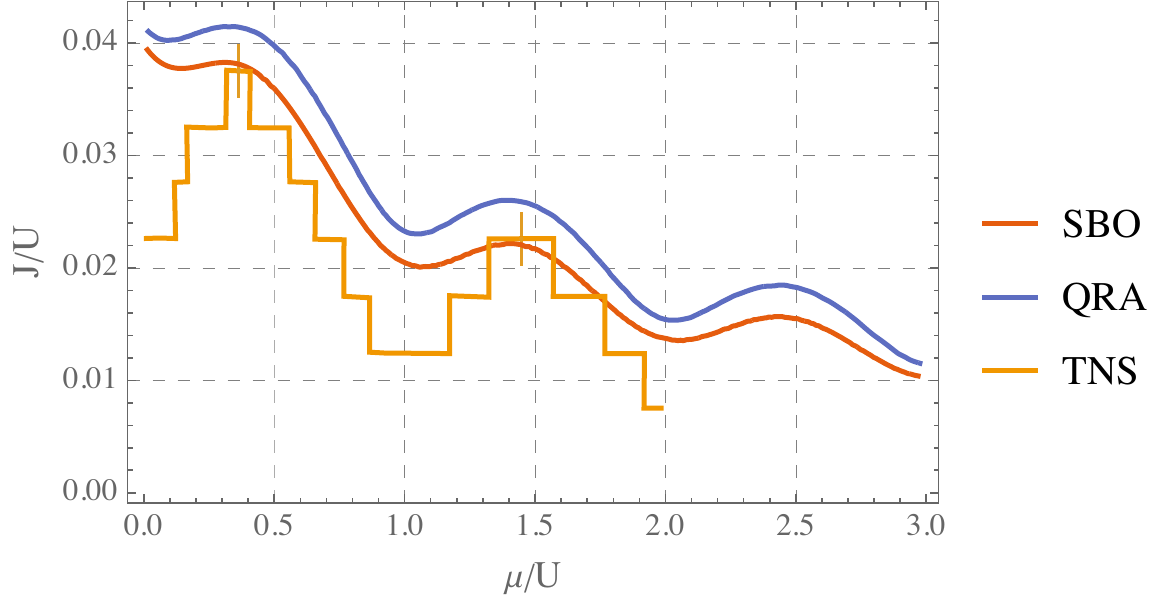}\caption{Comparison of the critical line at $T/U=0.1$ for three completely
different methods: SBO, QRA (Quantum Rotor Approach \citep{polak2007quantum,martin2025finite})
and TNS (tensor networks \citep{jahromi2020thermal}). \label{SBO_vs_others}}
\end{figure}

Finally, an interesting comparison is given in Fig. \ref{SBO_vs_others}.
Two completely different methods, namely Quantum Rotor Approach (QRA)
\citep{polak2007quantum,martin2025finite} and tensor networks (TNS)
\citep{jahromi2020thermal} are confronted with SBO. The QRA curve
lies substantially higher than the SBO one, while the prediction of
TNS is located even lower. The rough character of the last critical
line follows from the fact, that there was no dedicated algorithm
done in \citep{jahromi2020thermal} for drawing the phase boundary.
Rather, for a grid of possible $\mu/U$ and $J/U$ values, superfluid
density was determined. Critical line separates the zero superfluid
density region from that with its nonzero values. Error bars placed
at the tips of the lobes reflect the vertical step size of the mesh
and thus should be interpreted as maximal uncertainty. TNS study used
a maximal boson number per site of $n_{\mathrm{oc}}=2$, which corresponds
to the TSA for the first lobe. It makes the inter-lobe portion of
the critical line unreliable, as well as the entire second lobe. However,
tip of the first lobe should be accurate enough, as TSA doesn't modify
it that much (see Fig. \ref{fig:Comparison-of-the-T}). While the
SBO curve is closer to the predictions of TNS than the curve of QRA,
gathered observations are not enough to decide which method gives
better results.

An important advantage of the present approach is that all discussed
observables originate from one self-consistent Green-function calculation.
Phase boundaries, excitation spectra, momentum distributions and critical
temperatures therefore constitute mutually consistent predictions
rather than independent calculations adjusted for individual observables.

\section{SBO for two-component Bose-Hubbard model\label{sec:SBO-for-two-component}}

Following the success of the SBO method for single-species Bose-Hubbard
model, it would be good to extend it to the two-component case. However,
it is not obvious how it could account for the abundance of phases
present in the two-component case. Therefore, we start by generalizing
fundamental formulas of SBO in a way pursued by Rożek \citep{rozekmscthesis2016},
in particular assuming unbroken translational and $U\left(1\right)\times U\left(1\right)$
symmetry to avoid overcomplicating due to nonzero superfluid density
(of any kind).

In this approach, the atom-type degree of freedom is treated on equal
footing with position (site index). Thus SBO are of the form $\hat{L}_{\alpha\beta}^{\xi;i}$,
where $\xi=A,B$ and $\alpha,\beta$ are numbers of boson $\xi$.
Then the Hamiltonian acquires a form very similar to Eq. (\ref{SBO H}):
\begin{align}
\hat{H} & =\sum_{\xi i\alpha}E_{\alpha}^{\xi}\hat{L}_{\alpha\alpha}^{\xi;i}-\sum_{\left\langle i,j\right\rangle }\sum_{\xi\alpha\alpha^{\prime}\beta\beta^{\prime}}J_{\xi}T_{\alpha\alpha^{\prime}\beta\beta^{\prime}}\hat{L}_{\alpha\alpha^{\prime}}^{\xi;i}\hat{L}_{\beta\beta^{\prime}}^{\xi;j}\nonumber \\
 & +U_{AB}\sum_{i\alpha\beta}\alpha\beta\hat{L}_{\alpha\alpha}^{A;i}\hat{L}_{\beta\beta}^{B;i},\label{SBO_H2}
\end{align}
where
\begin{equation}
E_{\alpha}^{\xi}=\frac{U_{\xi}}{2}\alpha\left(\alpha-1\right)-\mu_{\xi}\alpha
\end{equation}
and $T_{\alpha\alpha^{\prime}\beta\beta^{\prime}}$ is given as before
by Eq. (\ref{SBO H}). Let us calculate a commutator of $\hat{L}_{\alpha\beta}^{\xi;i}$
with the Hamiltonian:
\begin{align}
 & \left[\hat{L}_{\alpha\beta}^{\xi;i},\hat{H}\right]=\nonumber \\
 & =\left[\hat{L}_{\alpha\beta}^{\xi;i},\hat{H}_{\xi}\right]+U_{AB}\left(\sum_{\beta^{\prime}}\beta^{\prime}\hat{L}_{\beta^{\prime}\beta^{\prime}}^{\zeta;i}\right)\left(\beta-\alpha\right)\hat{L}_{\alpha\beta}^{\xi;i},\label{comm2}
\end{align}
where $\zeta$ stands for the opposite boson type compared to $\xi$
and $\hat{H}_{\xi}$ is the Hamiltonian part containing only operators
acting on bosons $\xi$. There is no need to perform the remaining
calculations from scratch. It is sufficient to repeat the previous
ones (Sec. \ref{sec:Self-consistency-equations}) and trace the new
term (the second one in Eq. (\ref{comm2})). We focus on Green's functions
\begin{equation}
G_{\alpha\alpha^{\prime}\beta\beta^{\prime}}^{\xi;ij\pm}\left(t\right)=\mp\mathrm{i}\theta\left(\pm t\right)\left\langle \left[\hat{L}_{\alpha\alpha^{\prime}}^{\xi;i}\left(t\right),\hat{L}_{\beta\beta^{\prime}}^{\xi;j}\left(0\right)\right]\right\rangle ,
\end{equation}
which contain only the same type of SBO. Since 
\begin{equation}
-G_{\left[H,L_{\alpha\alpha^{\prime}}^{\xi;i}\right],L_{\beta\beta^{\prime}}^{\xi;j}}\left(\omega\right)=G_{\left[L_{\alpha\alpha^{\prime}}^{\xi;i},H\right],L_{\beta\beta^{\prime}}^{\xi;j}}\left(\omega\right),
\end{equation}
the right-hand-side of Eq. (\ref{EoM_step}) is extended by an additive
term
\begin{equation}
G_{U_{AB}\left(\sum_{\gamma}\gamma\hat{L}_{\gamma\gamma}^{\zeta;i}\right)\left(\alpha^{\prime}-\alpha\right)\hat{L}_{\alpha\alpha^{\prime}}^{\xi;i},L_{\beta\beta^{\prime}}^{\xi;j}}\left(\omega\right).\label{term}
\end{equation}
RPA decoupling reduces expression (\ref{term}) to
\begin{equation}
U_{AB}\left\langle \sum_{\gamma}\gamma\hat{L}_{\gamma\gamma}^{\zeta}\right\rangle \left(\alpha^{\prime}-\alpha\right)G_{\bm{\alpha\beta}}^{\xi;ij}\left(\omega\right).\label{term2}
\end{equation}
After Fourier transforming to $k$-space, formula (\ref{term2}) can
be moved to the left-hand-side of Eq. (\ref{EoM_step4}) to modify
$\left(\omega+E_{\alpha}^{\xi}-E_{\alpha^{\prime}}^{\xi}\right)G_{\bm{\alpha\beta}}^{\xi}\left(\omega,k\right)$
to
\begin{equation}
\left(\omega+\tilde{E}_{\alpha}^{\xi}-\tilde{E}_{\alpha^{\prime}}^{\xi}\right)G_{\bm{\alpha\beta}}^{\xi}\left(\omega,k\right),
\end{equation}
where
\begin{equation}
\tilde{E}_{\alpha}^{\xi}=\frac{U_{\xi}}{2}\alpha\left(\alpha-1\right)-\left(\mu_{\xi}-U_{AB}\left\langle \sum_{\gamma}\gamma\hat{L}_{\gamma\gamma}^{\zeta}\right\rangle \right)\alpha.\label{mu_renorm}
\end{equation}

This result is the end of calculations needed to be done here, the
rest is interpretation. Expectation value in Eq. (\ref{mu_renorm})
is just the mean number of bosons $\zeta$ (on each site):
\begin{equation}
\left\langle \sum_{\gamma}\gamma\hat{L}_{\gamma\gamma}^{\zeta}\right\rangle =\left\langle n_{\zeta}\right\rangle .
\end{equation}
Therefore, the equations governing given bosonic type are just identical
to those obtained before, but with a renormalized chemical potential
via the formula
\begin{equation}
\mu_{\xi}^{\prime}=\mu_{\xi}-U_{AB}\left\langle n_{\zeta}\right\rangle .\label{reno}
\end{equation}

It is exactly what is obtained, if one treats Hamiltonian (\ref{SBO_H2})
as two separate Bose-Hubbard models, with density-density coupling
between them treated in a mean-field manner. No new phases and generally
no new behavior can arise in this theory beyond what is known from
the single-species Bose-Hubbard model. Moreover, if boson densities
$n_{A}$ and $n_{B}$ are treated as external parameters, to which
chemical potentials $\mu_{A}$ and $\mu_{B}$ adjust, this SBO development
predicts literally nothing -- i. e. no effect of one species on the
other. Nevertheless, this negative result carries important methodological
information. It demonstrates that the inability to reproduce inter-species
quantum fluctuations does not originate from the standard basis operator
representation itself but rather from the RPA decoupling employed
to close the hierarchy of Green’s functions. Consequently, future
developments should primarily focus on improving the decoupling scheme
instead of modifying the operator basis.

However, nontrivial predictions can be obtained, if one wants to construct
the critical line in terms of chemical potentials as free parameters.
Of course, their value is limited, because they are just mean-field
results, but they give an idea how peculiar distortion is caused by
such a straightforward effect as crowding out of one bosonic species
by the other from the lattice.

\section{Results for two-component Bose-Hubbard model\label{sec:Results-for-two-component}}

\begin{figure*}
\centering{}\includegraphics[scale=0.5]{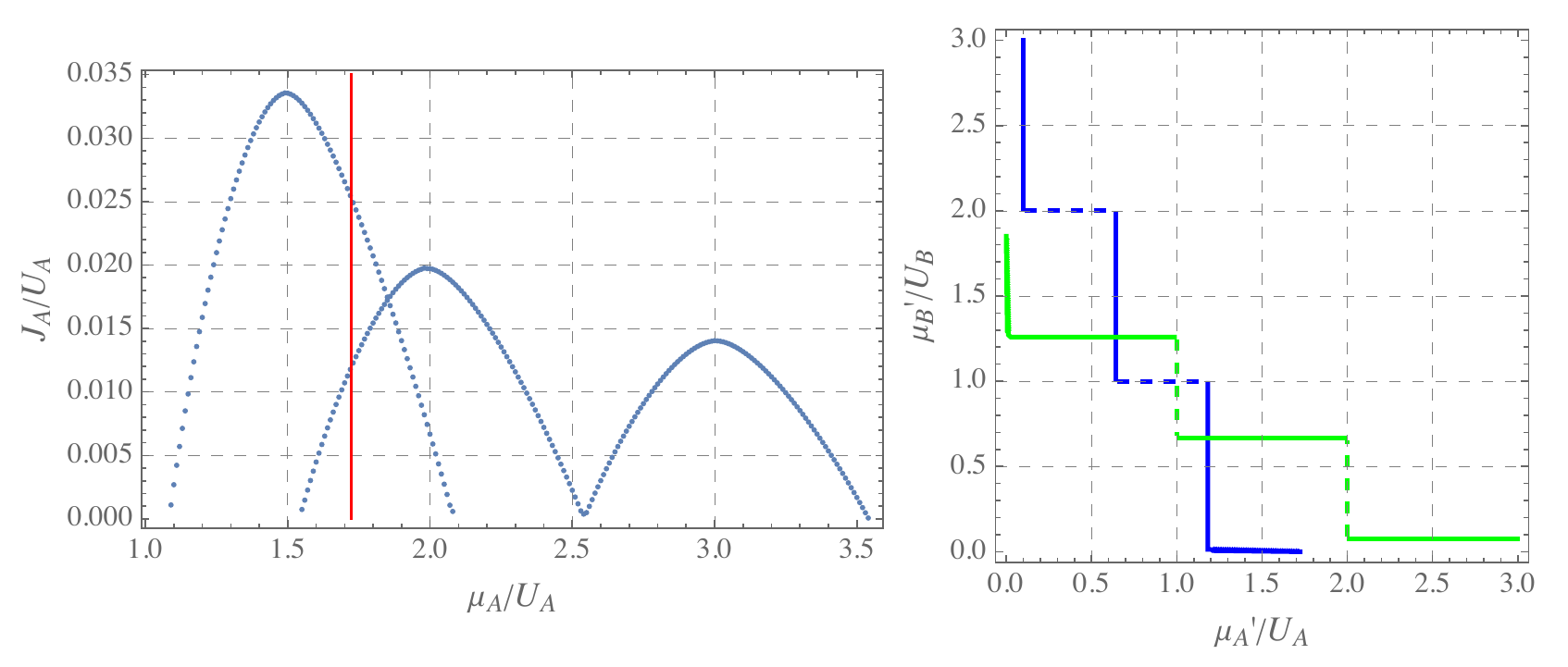}\caption{Overlapping of lobes ($T=0$) caused by mixing two bosonic species
(left) and a graphical representation of the self-consistency equations
for the point indicated by the red vertical line (right).\label{fig:Overlapping-of-lobes}}
\end{figure*}

Since we deal with two separate Bose-Hubbard models with mutual coupling
treated in mean-field manner, it is good to have pre-computed results
regarding a single Bose-Hubbard model alone (Sec. \ref{sec:Results-for-single-component}).
Each of them is evaluated for chemical potential $\mu_{\xi}^{\prime}$,
which is calculated self-consistently from Eq. (\ref{reno}). This
creates a system of two equations. However, not every solution is
a stable one. Assuming that the renormalized chemical potential is
attracted towards the value it should take on the basis of Eq. (\ref{reno}),
a phenomenological differential equation for $\mu_{\xi}^{\prime}$
can be written as
\begin{equation}
\frac{\partial}{\partial t}\mu_{\xi}^{\prime}=\mu_{\xi}-U_{AB}\left\langle n_{\zeta}\right\rangle -\mu_{\xi}^{\prime}.\label{diff1}
\end{equation}
Writing it in a matrix form and expanding the right-hand-side around
a solution $\left(\mu_{A\text{sol}}^{\prime},\mu_{B\text{sol}}^{\prime}\right)$
gives
\begin{align}
\frac{\partial}{\partial t}\begin{bmatrix}\mu_{A}^{\prime}\\
\mu_{B}^{\prime}
\end{bmatrix} & =-\begin{bmatrix}1 & U_{AB}\frac{\partial\left\langle n_{B}\right\rangle \left(\mu_{B}^{\prime}\right)}{\partial\mu_{B}^{\prime}}\\
U_{AB}\frac{\partial\left\langle n_{A}\right\rangle \left(\mu_{A}^{\prime}\right)}{\partial\mu_{A}^{\prime}} & 1
\end{bmatrix}\nonumber \\
 & \times\begin{bmatrix}\mu_{A}^{\prime}-\mu_{A\text{sol}}^{\prime}\\
\mu_{B}^{\prime}-\mu_{B\text{sol}}^{\prime}
\end{bmatrix}.\label{diff2}
\end{align}
For a stable solution, matrix in Eq. (\ref{diff2}) needs eigenvalues
with positive real part. This produces a condition that
\begin{equation}
\left|U_{AB}\right|\sqrt{\frac{\partial\left\langle n_{A}\right\rangle \left(\mu_{A}^{\prime}\right)}{\partial\mu_{A}^{\prime}}\frac{\partial\left\langle n_{B}\right\rangle \left(\mu_{B}^{\prime}\right)}{\partial\mu_{B}^{\prime}}}<1.\label{cond}
\end{equation}
Presented model of stability is definitely a toy model (Eq. (\ref{cond})
is not the precise condition), but it allows to identify stable and
unstable solutions in Fig. \ref{fig:Overlapping-of-lobes} ($T=0$).
The middle solution is unstable, because derivatives under the square
root are very large (actually divergent). Solutions on the ends have
these derivatives vanishing.

Presence of multiple solutions leads to overlapping of the lobes.
Such situation should be understood as coexistence of many branches
and the system follows that one in which it happened to be prepared.
Jumping to another branch requires ending of the previous. Thus, if
chemical potential is treated as the main parameter, first order transitions
occur. None, two, but also three lobes can overlap depending on the
chosen parameters. Figure \ref{fig:Overlapping-of-lobes} is plotted
for $U_{A}=1.2$, $U_{B}=1.1$, $U_{AB}=0.65$, $\mu_{B}/U_{B}=1.85$
and $T=0$.

\begin{figure*}
\centering{}\includegraphics[scale=0.5]{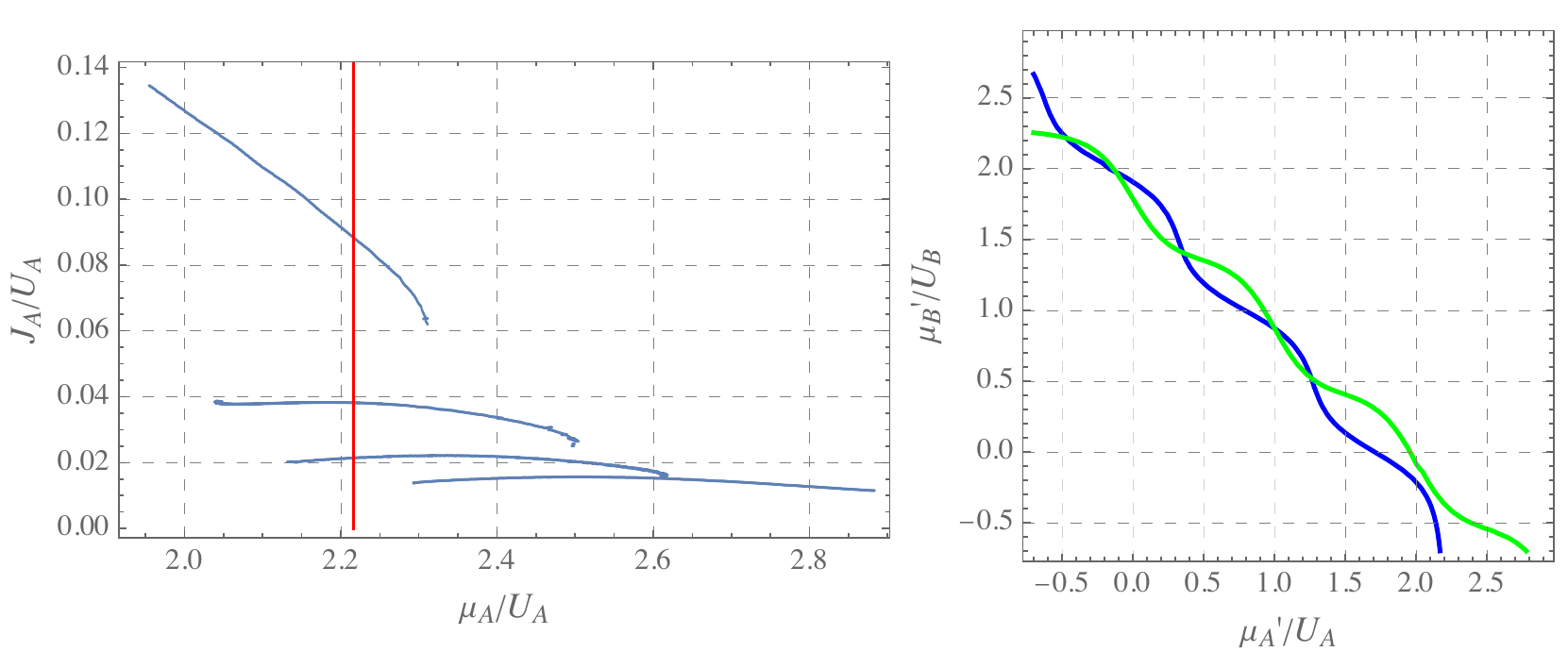}\caption{Overlapping lobes at nonzero temperature (left) and a graphical representation
of the self-consistency equations for the point indicated by the red
vertical line (right).\label{fig:Overlapping-lobes-at}}
\end{figure*}

Figure \ref{fig:Overlapping-lobes-at} shows the same phenomenon of
lobes overlapping, but at nonzero temperature $T_{A}/U_{A}=0.1$,
plotted for $U_{A}=U_{B}=0.96$, $U_{AB}=0.912$, $\mu_{B}/U_{B}=2.3$.
Now, even more solutions exist and a different argument to classify
them into stable and unstable ones can be given. Namely, each solution
corresponds to an extremum of free energy (since mean-field method
can be understood as a result of variational method). Some of them
are minima, which correspond to stable solutions, while others are
maxima, representing unstable ones. They must appear alternately,
with unstable solutions lying in the middle. Condition from Eq. (\ref{cond})
gives only a qualitative description of stability, after all, Eq.
(\ref{diff1}) was proposed heuristically.

A comment regarding values of hopping $J_{B}$ is needed. At zero
temperature it doesn't affect $\left\langle n_{B}\right\rangle $,
so any subcritical (or critical) $J_{B}$ can be taken. However, at
nonzero temperature $\left\langle n_{B}\right\rangle $ depends on
$J_{B}$ and some choice is needed. The one used here is to pick critical
$J_{B}$.

Regime, in which presented results have a chance to be reliable, corresponds
to subcritical hopping $J_{A}$ and $J_{B}$ (so that both species
are in the Mott phase) and $\left|U_{AB}\right|<\sqrt{U_{A}U_{B}}$,
so that no translational symmetry breaking (checkerboard or demixing)
can occur.

\section{Conclusion\label{sec:Conclusion}}

The enhanced SBO method for the single-component Bose-Hubbard model
fixes the following issues present in the previous implementation.
Discontinuities in the critical line ($J/U$ vs. $\mu/U$) are removed
and thus accuracy at nonzero temperatures and near-integer values
of $\mu/U$ is improved. Dependence of the critical temperature on
interaction strength ($T_{c}/J$ vs. $U/J$) is corrected for higher
temperatures, making the values consistent with experimental data
within its statistical error and also closer to the Monte-Carlo values.
Tensor network results\citep{jahromi2020thermal} are consistent with
those obtained by SBO method in the expected regime, while critical
line ($J/U$ vs. $\mu/U$) of Quantum Rotor Approach\citep{martin2025finite}
is systematically slightly above the previous two, but sharing the
same qualitative features.

An additional advantage of the present formulation is that the number
of retained local Fock states becomes a systematic convergence parameter.
This makes it possible to distinguish truncation effects from limitations
of the RPA decoupling. Beyond improving the phase diagram, the developed
formalism consistently describes several independent physical observables
within one Green-function framework, including excitation spectra
and momentum distributions.

Ambiguity in writing the self-consistency equations in the SBO method
comes from the fact that RPA violates multiplication rules of SBO.
Provisional solutions are possible, but they make the method increasingly
complex, while generating new challenges. This drawback makes treatment
of multi-component models difficult. This is shown on the example
of two-component Bose-Hubbard model, where only intra-species thermal
and quantum fluctuations can be taken into account. Still, interesting
deformation of the phase diagrams (overlapping of lobes) can be captured,
as well as first-order phase transitions steered by changes in the
chemical potential.

\appendix

\section{Critical line and numerical challenges\label{sec:Critical-line-and}}

Solutions to the final equations (Eqs. (\ref{final}) and (\ref{norm}))
were found numerically in Mathematica by the the procedure $\mathtt{FindRoot}$
\citep{reference.wolfram_2025_findroot}. For given parameters in
the Hamiltonian $J$, $U$, $\mu$ and temperature $T$, the solution
is found very quickly, but a significant slow down is encountered
as hopping $t$ is increased. An issue common to a vast number of
methods, namely critical slowing down, appears to be present here
as well. While the delay is barely noticeable in the three-states
approximation (TSA, having $\mathcal{N}-\mathcal{M}+1=3$), it is
problematic for higher number of states. Understanding its origin
and fixing it requires examining behavior of Eq. (\ref{final}) near
the critical point.

A Green's function encodes excitation energies in its poles. Solution
for $G_{\bm{\alpha}_{1}\bm{\beta}}$ in Eq. (\ref{Gsol}) implies
that these poles come from inverting matrix $M\left(\omega,k\right)$.
Since
\begin{equation}
M\left(\omega,k\right)=M\left(0,k\right)+\omega I,
\end{equation}
where $I$ is the identity matrix, singular frequencies $\omega_{n}$
(appearing in Eq. (\ref{X})) correspond to minus eigenvalues of $M\left(0,k\right)$.
This way, for each $k$ the dispersion relation of excitations (quasiparticles)
can be generated. Number of obtained branches is exactly the number
of included states per site minus one (which is the size of matrix
$M$).

The phase transition from Mott insulator to superfluid is associated
with mode softening, i. e. there appears a particle or hole excitation
that costs no energy. In other words, matrix $M\left(0,k\right)$
becomes singular for some value $k$ and as can be observed, it is
$k=0$. When the system is very close to the phase transition and
$k=0$ (or precisely at the critical point and $k\rightarrow0$),
$\omega_{n}$ from Eq. (\ref{X}) approaches zero and $f\left(\omega_{n}\right)$
diverges:
\begin{equation}
f\left(\omega_{n}\right)=\frac{1}{e^{\omega_{n}/T}-1}\cong\frac{T}{\omega_{n}}.\label{fasym}
\end{equation}
$k=0$ corresponds to $x=3$ in Eq. (\ref{final}). Thus $\rho_{s}$
tends to zero and elements of matrix $X$ diverge. This competition
leads to large numerical errors in the evaluation of final equations
and badly affects convergence of the root finding algorithm. To circumvent
this issue, asymptotic behavior of the integrand in Eq. (\ref{final})
near $x=3$ at the critical point has to be described and used to
avoid multiplication of small and large numbers. First, the density
of states $\rho_{s}\left(x\right)$ can be approximated as follows
for $x\rightarrow3^{-}$:
\begin{align}
 & \rho_{s}\left(x\right)=\frac{1}{\left(2\pi\right)^{3}}\int\mathrm{d}^{3}k\,\delta\left(x-\frac{1}{2}\epsilon\left(k\right)\right)\nonumber \\
 & \cong-\frac{\partial}{\partial x}\frac{1}{\left(2\pi\right)^{3}}\int\mathrm{d}^{3}k\,\theta\left(3-x-\left(\frac{1}{2}k_{x}^{2}+\frac{1}{2}k_{y}^{2}+\frac{1}{2}k_{z}^{2}\right)\right)\nonumber \\
 & =\frac{1}{\sqrt{2}\pi^{2}}\left(3-x\right)^{1/2}.\label{rho_asymp}
\end{align}

There are two qualitatively different situations. The first corresponds
to particle-hole symmetry (so-called tip of the lobe due to its position
on the critical line at zero temperature). In consequence, there are
two zero eigenvalues $\omega_{-1}$ and $\omega_{1}$, which come
from different branches at $k=0$. The other situation lacks the aforementioned
symmetry, so there is only one zero eigenvalue $\omega_{0}$ at $k=0$.

First, we focus on the particle-hole symmetric case. In order to handle
divergence of the $M$ matrix, it is convenient to write its inverse
as
\begin{equation}
M^{-1}\left(\omega,\epsilon\right)=\frac{\mathrm{adj}M\left(\omega,\epsilon\right)}{\prod_{n}\left[\omega-\omega_{n}\left(\epsilon\right)\right]},
\end{equation}
where $\mathrm{adj}$ stands for an adjugate of a matrix and dependence
on the occupancies $D$ is omitted in notation. Then, using additionally
Eq. (\ref{fasym}), matrix $X$ can be approximated for $\epsilon\rightarrow6^{-}$
($x=\epsilon/2\rightarrow3^{-}$) as
\begin{align}
 & X_{\alpha\beta}\left(\epsilon\right)\cong-\frac{T\left[\mathrm{adj}M\left(0,\epsilon\right)\right]_{\alpha\beta}}{\omega_{1}\left(\epsilon\right)\omega_{-1}\left(\epsilon\right)\prod_{m\neq\pm1}\left[-\omega_{m}\left(\epsilon\right)\right]}\nonumber \\
 & =-\frac{T\left[\mathrm{adj}M\left(0,6\right)\right]_{\alpha\beta}}{\det M\left(0,\epsilon\right)}.\label{Xasym}
\end{align}
Critical point at the tip of the lobe is characterized by a Dirac
cone in its dispersion centered at zero energy and $k=0$. Thus
\begin{equation}
\omega_{\pm1}\left(k\right)\cong\pm\mathrm{const.}\times\left|k\right|
\end{equation}
and
\begin{equation}
\det M\left(0,\epsilon\left(k\right)\right)\cong\mathrm{const.}\times k^{2}.\label{detM}
\end{equation}
Using similar expansion as in Eq. (\ref{rho_asymp}), the following
relation holds for $\epsilon\rightarrow6^{-}$:
\begin{equation}
\left|k\right|\cong\sqrt{6-\epsilon}=\sqrt{2\left(3-x\right)}.\label{k}
\end{equation}
Combining Eqs. (\ref{detM}) and (\ref{k}) gives
\begin{equation}
\det M\left(0,\epsilon\left(k\right)\right)\cong-\left.\frac{\partial}{\partial\epsilon}\det M\left(0,\epsilon\right)\right|_{\epsilon=6}\times k^{2}.\label{detM2}
\end{equation}
Plugging Eqs. (\ref{detM2}) and (\ref{k}) to Eq. (\ref{Xasym})
leads to
\begin{equation}
X_{\alpha\beta}\left(\epsilon\right)\cong\frac{T\left[\mathrm{adj}M\left(0,6\right)\right]_{\alpha\beta}}{\left.\frac{\partial}{\partial\epsilon}\det M\left(0,\epsilon\right)\right|_{\epsilon=6}}\frac{1}{6-\epsilon}.\label{Xasym2}
\end{equation}
The integrand of Eq. (\ref{final}) near $x=3$ behaves as
\begin{align}
 & \rho_{s}\left(x\right)X_{\alpha-1,\beta-1}\left(2x\right)\cong\frac{T}{2\sqrt{2}\pi^{2}}\nonumber \\
 & \times\frac{\left[\mathrm{adj}M\left(0,6\right)\right]_{\alpha-1,\beta-1}}{\left.\frac{\partial}{\partial\epsilon}\det M\left(0,\epsilon\right)\right|_{\epsilon=6}}\frac{1}{\sqrt{3-x}},
\end{align}
so it has an integrable singularity. The integral should be performed
analytically in some small interval $\left(x^{*},3\right)$, where
the approximation is highly accurate. Numerical integration was less
stable in this interval as it is typical for singular integrands.
The modified final equations become
\begin{align}
 & \sqrt{\beta}D_{\beta}=\nonumber \\
 & =\left(D_{\beta-1}-D_{\beta}\right)\sum_{\alpha}\sqrt{\alpha}\int_{-d}^{d}\mathrm{d}x\,\rho_{s}\left(x\right)X_{\alpha-1,\beta-1}\left(2x,D\right)\nonumber \\
 & +\left(D_{\beta-1}-D_{\beta}\right)\frac{T\sqrt{3-x^{*}}}{\sqrt{2}\pi^{2}\left.\frac{\partial}{\partial\epsilon}\det M\left(0,\epsilon,D\right)\right|_{\epsilon=6}}\nonumber \\
 & \times\sum_{\alpha}\sqrt{\alpha}\left[\mathrm{adj}M\left(0,6,D\right)\right]_{\alpha-1,\beta-1}.\label{final_corr}
\end{align}
Value of $x^{*}$ was chosen between $2.94$ and $2.99$. For computing
a derivative of a determinant, the Jacobi's formula \citep{magnus2019matrix}
can be used:
\begin{align}
 & \left.\frac{\partial}{\partial\epsilon}\det M\left(0,\epsilon,D\right)\right|_{\epsilon=6}=\nonumber \\
 & =\mathrm{Tr}\left[\mathrm{adj}M\left(0,6,D\right)\left.\frac{\partial}{\partial\epsilon}M\left(0,\epsilon,D\right)\right|_{\epsilon=6}\right],
\end{align}
where
\begin{equation}
\frac{\partial}{\partial\epsilon}M_{\alpha\alpha_{1}}\left(\omega,\epsilon,D\right)=J\sqrt{\left(\alpha+1\right)\left(\alpha_{1}+1\right)}\left(D_{\alpha}-D_{\alpha+1}\right).
\end{equation}

Now, we focus on the situation without particle-hole symmetry. Although
the derivation in this case is slightly different, the final result
is identical to Eq. (\ref{final_corr}). Here, only one eigenvalue
$\omega_{0}$ tends to zero as $k\rightarrow0$ and $\omega_{0}\sim k^{2}$.
Analogously to Eq. (\ref{Xasym}) matrix $X$ can be approximated
as
\begin{align}
 & X_{\alpha\beta}\left(\epsilon\right)\cong\frac{T}{\omega_{0}\left(\epsilon\right)}\frac{\left[\mathrm{adj}M\left(0,6\right)\right]_{\alpha\beta}}{\prod_{m\neq0}\left[-\omega_{m}\left(6\right)\right]}\nonumber \\
 & =-\frac{T\left[\mathrm{adj}M\left(0,6\right)\right]_{\alpha\beta}}{\det M\left(0,\epsilon\right)},\label{XasymNew}
\end{align}
which is identical to (\ref{Xasym}). Also Eq. (\ref{detM}) is valid,
but now the $k^{2}$ proportionality comes from a single eigenvalue,
not from two linearly scaling with $\left|k\right|$. The rest of
the derivation is thus identical to that of the previous section.

It is a very convenient situation, that a single modification (Eq.
(\ref{final_corr})) to the final equations suffices to eliminate
numerical instability at the critical point. Additionally, at zero
temperature the correction vanishes. It is a manifestation of the
fact that the Bose-Einstein distribution can be approximated as
\begin{equation}
f\left(\omega\right)=\begin{cases}
-1 & \text{for }\omega<0\\
0 & \text{for }\omega>0
\end{cases},\;\text{at }T=0,
\end{equation}
which is free of any divergences. Unfortunately, there is an additional
bad behavior of the final equations at nonzero temperatures which
happens when transitioning between lobes, i. e. for integer values
of $\mu/U$. Then, matrix $M\left(0,k\right)$ becomes singular, because
both $E_{\alpha}-E_{\alpha+1}$ and $D_{\alpha}-D_{\alpha+1}$ vanish
for $\alpha=\mu/U$. Vanishing of $D_{\alpha}-D_{\alpha+1}$ is probably
less obvious, but can be seen from the numerical calculations. Then
one row of matrix $M\left(0,k\right)$ becomes only zeros, which implies
$\det M\left(0,k\right)=0$ for any $k$. It means that one band in
the dispersion is entirely flat and has zero energy. This exact point
cannot be captured even by Eq. (\ref{final_corr}). However, it can
be approached very closely from both sides on the $\mu/U$ axis.

\section{Conceptual problems with the SBO method\label{sec:Conceptual-problems-with}}

The main controversy arising in the SBO method comes from Eq. (\ref{sc-eq}).
It can be written in many nonequivalent ways. Any matrix $c_{\beta\alpha}$
with nonzero diagonal elements does the job:
\begin{align}
 & c_{\beta\beta}D_{\beta}=\left\langle \hat{L}_{\beta,\beta-1}\left(\sum_{\alpha}c_{\beta\alpha}\hat{L}_{\alpha-1,\alpha}\right)\right\rangle \nonumber \\
 & =\sum_{\alpha}c_{\beta\alpha}\left\langle \hat{L}_{\beta,\beta-1}\hat{L}_{\alpha-1,\alpha}\right\rangle \nonumber \\
 & =\sum_{\alpha}\frac{c_{\beta\alpha}}{N}\sum_{k}\sum_{n}f\left(\omega_{n}\right)\mathrm{Res}_{\omega=\omega_{n}}\left\{ G_{\alpha-1,\alpha,\beta,\beta-1}\left(\omega,k\right)\right\} .
\end{align}
The choice of $c_{\beta\alpha}$ used in Sec. \ref{sec:Self-consistency-equations}
is somewhat special, because it produces ``the most natural'' linear
combination of SBO. In other words, only for $c_{\beta\alpha}=c_{\beta}\sqrt{\alpha}$
(with any sequence $c_{\beta}$ with nonzero elements) we obtain an
annihilation operator from $\sum_{\alpha}c_{\beta\alpha}\hat{L}_{\alpha-1,\alpha}$,
which enters the original Hamiltonian unlike single SBO. However,
this is a very hand-wavy argument. It has been explicitly checked
in \citep{sajna2015ground} that using a different straightforward
setting, namely $c_{\beta\alpha}=\delta_{\beta\alpha}$, gives worse
results in terms of the height of zero-temperature lobes. They are
lower than Monte-Carlo curves, but still improving mean-field. The
reason behind ambiguity in choosing self-consistency equations comes
from the fact that there are actually more equations than unknowns.
The required relations come from multiplication rules for SBO:
\begin{equation}
\left\langle \hat{L}_{\beta,\beta-1}\hat{L}_{\alpha-1,\alpha}\right\rangle =D_{\beta}\delta_{\alpha\beta},
\end{equation}
which gives $\left(\mathcal{N}-\mathcal{M}\right)^{2}+1$ equations
(the sum rule $\sum_{\alpha}D_{\alpha}=1$ is added) for $\mathcal{N}-\mathcal{M}+1$
unknowns. This system should be consistent, provided that the Green's
function is exact. However, if choosing a different linear combination
of the available equations leads to different results, the system
is inconsistent. Thus the RPA approximation violates in some way the
algebra of SBO operators.

In order to see the source of this violation, essence of the RPA approximation
should be exposed. Its formulation in Eq. (\ref{RPA}) reminds of
an exact identity regarding commutators
\begin{equation}
\left[\hat{A}\hat{B},\hat{C}\right]=\hat{A}\left[\hat{B},\hat{C}\right]+\left[\hat{A},\hat{C}\right]\hat{B},
\end{equation}
but a mean-field-like decoupling of the right-hand-side averages is
done. However, it should be emphasized that the RPA decoupling is
beyond mean-field approximation. Let us write Eq. (\ref{RPA}) in
a different form. First, by $\delta\hat{A}=\hat{A}-\left\langle \hat{A}\right\rangle $
a deviation of any operator $\hat{A}$ from its expectation value
is denoted. Moving $\left\langle \hat{A}\right\rangle \left\langle \left[\hat{B},\hat{C}\right]\right\rangle $
to the left-hand-side of Eq. (\ref{RPA}) gives
\begin{equation}
\left\langle \left[\delta\hat{A}\hat{B},\hat{C}\right]\right\rangle \overset{\mathrm{RPA}}{\approx}\left\langle \left[\hat{A},\hat{C}\right]\right\rangle \left\langle \hat{B}\right\rangle .
\end{equation}
Since a constant commutes with everything $\left\langle \left[\hat{A},\hat{C}\right]\right\rangle =\left\langle \left[\delta\hat{A},\hat{C}\right]\right\rangle $.
Moving this term to the opposite site yields
\begin{equation}
\left\langle \left[\delta\hat{A}\delta\hat{B},\hat{C}\right]\right\rangle \overset{\mathrm{RPA}}{\approx}0.
\end{equation}
Noting again that a constant commutes with everything and expanding
the commutator leads to
\begin{equation}
\left\langle \delta\hat{A}\delta\hat{B}\delta\hat{C}\right\rangle \overset{\mathrm{RPA}}{\approx}\left\langle \delta\hat{C}\delta\hat{A}\delta\hat{B}\right\rangle .\label{RPA_alt}
\end{equation}
Such formulation of the RPA approximation can be formulated in one
sentence: A three-operator connected correlation function can be cyclically
permuted. Equivalently $\left\langle \left[\delta\hat{A}\delta\hat{B},\delta\hat{C}\right]\right\rangle =0$.
This approximation scheme is more sophisticated than mean-field, which
prescribes $\delta\hat{A}\delta\hat{B}\overset{\mathrm{mf}}{\approx}0$.

Similarly to any decoupling procedure, RPA should be applied only
to specific operators $\hat{A}$, $\hat{B}$, $\hat{C}$, not just
any triple. For example, mean-field approximation is applied to operators
pairs which act on different lattice sites. Product of two operators
acting on the same site is just one resulting operator acting on this
same site, which should not be decoupled. In the SBO method situation
is more complicated, because working in the Heisenberg picture, time
$t$ enters the equations. A natural rule would be to decouple $\left\langle \delta\hat{A}\delta\hat{B}\delta\hat{C}\right\rangle $
provided that operators $\hat{A}$, $\hat{B}$ and $\hat{C}$ correspond
to different space-time points. It is problematic, because time is
continuous. However, certainly, if $\hat{B}$ and $\hat{C}$ are SBO
and act on the same site being evaluated at the same time, no decoupling
should be applied, because $\left\langle \delta\hat{A}\delta\hat{B}\delta\hat{C}\right\rangle $
is reducible to a two-operator correlation function. This requirement
is breached by the way Sec. \ref{sec:Self-consistency-equations}
uses RPA, which violates the SBO algebra and produces discussed ambiguities.
Let us show it explicitly. A real-time version of Eq. (\ref{decoupling})
can be written as
\begin{equation}
\bm{G}_{\bm{\mu}\bm{\alpha}_{1}\bm{\beta}}^{ij\pm}\left(t\right)\overset{\mathrm{RPA}}{\cong}\left\langle \hat{L}_{\bm{\mu}}\right\rangle \sum_{r\text{n.t.}i}G_{\bm{\alpha}_{1}\bm{\beta}}^{rj\pm}\left(t\right).\label{decoup}
\end{equation}
Examining it for $t=0^{\pm}$ exactly leads to
\begin{align}
 & \bm{G}_{\bm{\mu}\bm{\alpha}_{1}\bm{\beta}}^{ij\pm}\left(0^{\pm}\right)=\nonumber \\
 & =\mp\mathrm{i}\sum_{r\text{n.t.}i}\left\langle \left[\hat{L}_{\bm{\mu}}^{i}\hat{L}_{\bm{\alpha}_{1}}^{r},\hat{L}_{\bm{\beta}}^{j}\right]\right\rangle \nonumber \\
 & =\mp\mathrm{i}\sum_{r\text{n.t.}i}\left\langle \left[\hat{L}_{\bm{\mu}}^{i}\delta\hat{L}_{\bm{\alpha}_{1}}^{r},\delta\hat{L}_{\bm{\beta}}^{j}\right]\right\rangle \nonumber \\
 & =\mp\mathrm{i}\sum_{r\text{n.t.}i}\left\langle \left[\delta\hat{L}_{\bm{\mu}}^{i}\delta\hat{L}_{\bm{\alpha}_{1}}^{r},\delta\hat{L}_{\bm{\beta}}^{j}\right]\right\rangle \nonumber \\
 & +\left\langle \hat{L}_{\bm{\mu}}\right\rangle \sum_{r\text{n.t.}i}G_{\bm{\alpha}_{1}\bm{\beta}}^{rj\pm}\left(0^{\pm}\right).
\end{align}
RPA approximation neglects $\left\langle \left[\delta\hat{L}_{\bm{\mu}}^{i}\delta\hat{L}_{\bm{\alpha}_{1}}^{r},\delta\hat{L}_{\bm{\beta}}^{j}\right]\right\rangle $,
but for $j=i$ or $j$ lying next to $i$ it is inconsistent. For
$j=i$:
\begin{align}
 & \sum_{r\text{n.t.}i}\left\langle \left[\delta\hat{L}_{\bm{\mu}}^{i}\delta\hat{L}_{\bm{\alpha}_{1}}^{r},\delta\hat{L}_{\bm{\beta}}^{i}\right]\right\rangle =\nonumber \\
 & =\sum_{r\text{n.t.}i}\left\langle \left[\delta\hat{L}_{\bm{\mu}}^{i},\delta\hat{L}_{\bm{\beta}}^{i}\right]\delta\hat{L}_{\bm{\alpha}_{1}}^{r}\right\rangle \nonumber \\
 & =\sum_{r\text{n.t.}i}\left\langle \left[\hat{L}_{\bm{\mu}}^{i},\hat{L}_{\bm{\beta}}^{i}\right]\delta\hat{L}_{\bm{\alpha}_{1}}^{r}\right\rangle \nonumber \\
 & =\sum_{r\text{n.t.}i}\left\langle \left(\hat{L}_{\mu\beta^{\prime}}^{i}\delta_{\mu^{\prime}\beta}-\hat{L}_{\beta\mu^{\prime}}^{i}\delta_{\beta^{\prime}\mu}\right)\delta\hat{L}_{\bm{\alpha}_{1}}^{r}\right\rangle \label{j=00003Di}
\end{align}
and for $j$ neighboring with $i$:
\begin{align}
 & \sum_{r\text{n.t.}i}\left\langle \left[\delta\hat{L}_{\bm{\mu}}^{i}\delta\hat{L}_{\bm{\alpha}_{1}}^{r},\delta\hat{L}_{\bm{\beta}}^{j}\right]\right\rangle =\nonumber \\
 & =\left\langle \delta\hat{L}_{\bm{\mu}}^{i}\left[\hat{L}_{\bm{\alpha}_{1}}^{j},\hat{L}_{\bm{\beta}}^{j}\right]\right\rangle +\underset{r\neq j}{\sum_{r\text{n.t.}i}}\left\langle \left[\delta\hat{L}_{\bm{\mu}}^{i}\delta\hat{L}_{\bm{\alpha}_{1}}^{r},\delta\hat{L}_{\bm{\beta}}^{j}\right]\right\rangle \nonumber \\
 & =\left\langle \delta\hat{L}_{\bm{\mu}}^{i}\left(\hat{L}_{\alpha_{1}\beta^{\prime}}^{j}\delta_{\alpha_{1}^{\prime}\beta}-\hat{L}_{\beta\alpha_{1}^{\prime}}^{j}\delta_{\beta^{\prime}\alpha_{1}}\right)\right\rangle \nonumber \\
 & +\underset{r\neq j}{\sum_{r\text{n.t.}i}}\left\langle \left[\delta\hat{L}_{\bm{\mu}}^{i}\delta\hat{L}_{\bm{\alpha}_{1}}^{r},\delta\hat{L}_{\bm{\beta}}^{j}\right]\right\rangle .\label{r=00003Dj}
\end{align}
The last term in Eq. (\ref{r=00003Dj}) can be reasonably decoupled
to zero by RPA, but the first definitely not, similarly to the last
line in Eq. (\ref{j=00003Di}).

A natural way around this would be to extend Eq. (\ref{decoup}) by
necessary corrections needed when $i=j$ or $i$ and $j$ are neighbors.
However, they are applicable only at $t=0$. For large $t$ the corrections
should certainly be abandoned, since correlations decay over time,
but the crossover between small and large $t$ remains unknown. It
is possible to model this crossover by an exponential decay $e^{-t/\tau}$
with some characteristic time $\tau$. The complication this brings
is possible to handle, but it is difficult to identify self-consistency
equations involving $\tau$ and free of problems present before.

Maybe a different decoupling exists, which preserves the SBO algebra.
Unfortunately, we did not find a satisfying one. Even though results
for the Bose-Hubbard model are decently accurate, conceptually unclear
structure of the SBO method discourages from applying it to more complex
models.

\section{Using a single class of composite SBO\label{sec:Using-a-single}}

It can be argued that the disappointment brought by SBO applied to
Bose-Bose mixtures comes from its inappropriate usage. Introducing
two different ``species'' of SBO, namely $\hat{L}_{\alpha\beta}^{A;i}$
and $\hat{L}_{\alpha\beta}^{B;i}$ leads to their decoupling by RPA,
when they appear in a product. A more involved scheme of generalization,
and therefore more promising, consists in defining just a single class
of SBO $\hat{L}_{\alpha\beta}^{i}$. However, they are composite in
a sense that index $\alpha$ (and $\beta$ likewise) encodes full
occupancy of site $i$. In other words, $\alpha=\left(\alpha_{A},\alpha_{B}\right)$
is a double index. Then $\bm{\alpha}$ becomes a four-index, so the
Green's function $G_{\bm{\alpha\beta}}$ has actually eight indices.
In Sec. \ref{sec:Self-consistency-equations} only Green's function
based on $\hat{a}$, $\hat{a}^{\dag}$ were relevant, so restriction
to $\bm{\alpha}\in\uparrow$ and $\bm{\beta}\in\downarrow$ was done.
Here, choosing three different Green's functions based on operators
$\left(\hat{a},\hat{a}^{\dag}\right)$, $\left(\hat{b},\hat{b}^{\dag}\right)$
and (for example) $\left(\hat{a}\hat{b}^{\dag},\hat{a}^{\dag}\hat{b}\right)$
produces three classes of indices $\bm{\alpha}=\left(\alpha,\alpha^{\prime}\right)$:
\begin{enumerate}
\item Class $\mathscr{A}$:
\begin{equation}
\left|\alpha^{\prime}\right\rangle =\frac{1}{\sqrt{\alpha_{A}+1}}\hat{a}^{\dag}\left|\alpha\right\rangle .
\end{equation}
\item Class $\mathscr{B}$:
\begin{equation}
\left|\alpha^{\prime}\right\rangle =\frac{1}{\sqrt{\alpha_{B}+1}}\hat{b}^{\dag}\left|\alpha\right\rangle .
\end{equation}
\item Class $\mathscr{AB}$:
\begin{equation}
\left|\alpha^{\prime}\right\rangle =\frac{1}{\sqrt{\left(\alpha_{A}+1\right)\alpha_{B}}}\hat{a}^{\dag}\hat{b}\left|\alpha\right\rangle .
\end{equation}
\end{enumerate}
\vspace{-0.4\baselineskip}

Equation of motion can be written for each class:
\begin{align}
 & \sum_{\bm{\alpha}_{1}\in\mathscr{A}}\left[\left(\omega+E_{\alpha}-E_{\alpha^{\prime}}\right)\delta_{\bm{\alpha}\bm{\alpha}_{1}}\right.\nonumber \\
 & \left.+J_{A}\sqrt{\left(\alpha_{A}+1\right)\left(\alpha_{1A}+1\right)}\left(D_{\alpha}-D_{\alpha^{\prime}}\right)\epsilon\left(k\right)\right]\nonumber \\
 & \times G_{\bm{\alpha}_{1}\bm{\beta}}\left(\omega,k\right)=\left(D_{\alpha}-D_{\alpha^{\prime}}\right)\delta_{\alpha\beta^{\prime}}\delta_{\alpha^{\prime}\beta},\;\text{for }\bm{\alpha}\in\mathscr{A},\label{EoM_A}
\end{align}
\begin{align}
 & \sum_{\bm{\alpha}_{1}\in\mathscr{B}}\left[\left(\omega+E_{\alpha}-E_{\alpha^{\prime}}\right)\delta_{\bm{\alpha}\bm{\alpha}_{1}}\right.\nonumber \\
 & \left.+J_{B}\sqrt{\left(\alpha_{B}+1\right)\left(\alpha_{1B}+1\right)}\left(D_{\alpha}-D_{\alpha^{\prime}}\right)\epsilon\left(k\right)\right]\nonumber \\
 & \times G_{\bm{\alpha}_{1}\bm{\beta}}\left(\omega,k\right)=\left(D_{\alpha}-D_{\alpha^{\prime}}\right)\delta_{\alpha\beta^{\prime}}\delta_{\alpha^{\prime}\beta},\;\text{for }\bm{\alpha}\in\mathscr{B},\label{EoM_B}
\end{align}
\begin{align}
 & \left(\omega+E_{\alpha}-E_{\alpha^{\prime}}\right)G_{\bm{\alpha}\bm{\beta}}\left(\omega,k\right)\nonumber \\
 & =\left(D_{\alpha}-D_{\alpha^{\prime}}\right)\delta_{\alpha\beta^{\prime}}\delta_{\alpha^{\prime}\beta},\;\text{for }\bm{\alpha}\in\mathscr{AB}.\label{EoM_AB}
\end{align}
 Unfortunately, there are a few problems with them. First of all,
Eq. (\ref{EoM_AB}) predicts trivial poles for $G_{\bm{\alpha}\bm{\beta}}$
when $\bm{\alpha}\in\mathscr{AB}$. This happens, because there is
no term in the Hamiltonian simultaneously swapping bosons $a$ and
$b$ (it can emerge as a virtual process for large $U_{AB}$, but
SBO method doesn't know it). Moreover, Eq. (\ref{EoM_A}) can fix
all $D_{\left(\alpha_{A},\alpha_{B}\right)}$ given $D_{\left(0,\alpha_{B}\right)}$
for all $\alpha_{B}$, similarly to what happened in Sec. \ref{sec:Self-consistency-equations}.
Situation is analogous with Eq. (\ref{EoM_B}), which fixes all $D_{\left(\alpha_{A},\alpha_{B}\right)}$
given $D_{\left(\alpha_{A},0\right)}$ for all $\alpha_{A}$. This
is not very transparent and it is not obvious whether a reasonable
self-consistency equations can be formed.

As mentioned before in Sec. \ref{sec:Conceptual-problems-with}, conceptual
problems with the SBO method make it difficult to use in multi-component
cases. These observations together with Appendix \ref{sec:Conceptual-problems-with}
identify the RPA decoupling itself as the principal limitation of
the current formulation. Maybe some additional ideas are needed to
make it useful.

\end{document}